\documentclass[12pt,preprint]{aastex}

\shorttitle{Optical Monitoring of S5 0716+714}
\shortauthors{Wu et al.}

\received{}
\accepted{}

\begin{document}

\title{Optical Monitoring of BL Lac Object S5 0716+714 with High
       Temporal Resolution}

\author{Jianghua Wu}
 \affil{National Astronomical Observatories, Chinese Academy of
        Sciences, 20A Datun Road, Beijing 100012, China}
 \email{jhwu@bao.ac.cn}

\author{Bo Peng}
 \affil{National Astronomical Observatories, Chinese Academy of
        Sciences, 20A Datun Road, Beijing 100012, China}
 \email{pb@bao.ac.cn}

\author{Xu Zhou}
 \affil{National Astronomical Observatories, Chinese Academy of
        Sciences, 20A Datun Road, Beijing 100012, China}
 \email{zhouxu@bac.pku.edu.cn}

\author{Jun Ma}
 \affil{National Astronomical Observatories, Chinese Academy of
        Sciences, 20A Datun Road, Beijing 100012, China}

\author{Zhaoji Jiang}
 \affil{National Astronomical Observatories, Chinese Academy of
        Sciences, 20A Datun Road, Beijing 100012, China}
\and

\author{Jiansheng Chen}
 \affil{National Astronomical Observatories, Chinese Academy of
        Sciences, 20A Datun Road, Beijing 100012, China}

\begin{abstract}
Optical monitoring of S5 0716+714 was performed with a 60/90
Schmidt telescope in 2003 November and December and 2004 January for
studying the variability of the object on short timescales. Due to the
high brightness of the source we could carry out quasi-simultaneous
measurements in three bands with a temporal resolution of about 20
minutes by using one single telescope. Intraday and intranight variations
were observed with an overall change of $\sim$0.9 mag during the whole
campaign. Two outbursts were recorded on JD~2\,453\,005 and JD~2\,453\,009.
Minimum timescales of a few hours were derived from the light curves of
individual nights but were different from night to night. A
bluer-when-brighter chromatism was present when the object was in fast
flare, but was absent when it was in a relatively quiescent state. Our
results are basically consistent with the shock-in-jet model and demonstrate
that the geometrical effects can sometimes play an important role in the
variability of blazars.
\end{abstract}


\section{INTRODUCTION}
Blazars are a subset of active galactic nuclei (AGNs). These radio-loud
flat-spectrum objects exhibit the most rapid variations and the largest
amplitude variations among all AGNs. The variations are thought to
originate from a relativistic jet, which is believed to be oriented at a small
angle to our light of sight and which is probably powered and accelerated
by a rotating and accreting supermassive black hole. There are basically
two types of blazars, the BL Lac objects and the flat-spectrum radio quasars
(FSRQ), the former have a featureless optical continuum while the
latter show many strong and broad emission lines.

Variability studies of blazars have been essential
in understanding the physics of their central regions, which in general
cannot be resolved even with existing or planned optical/infrared
interferometers. The timescales, the spectral changes, and the
correlations and delays between variations in different continuum
components provide crucial information on the nature and location of
these components and on their interdependencies. These parameters can
be well studied with multi-frequency observational campaigns,
such as those coordinated for Mrk 421 \citep[e.g.,][]
{buckley96}, 3C 279 \citep[e.g.,][]{wehrle98}, S5 0716+714 \citep[e.g.,]
[]{raiteri03,wagner96}, and PKS 2155-304 \citep[e.g.,][]{urry97}.

The BL Lac object S5 0716+714 is well known for its intraday variability
(IDV) in the radio and optical bands and has been the target of many
monitoring programs, as mentioned above. \citet{nesci02} found a typical
variation rate of 0.02 mag per hour and a maximum rising rate of 0.16 mag
per hour for this object. \citet{heidt96} reported a period
of 4 day in the optical band while \citet{qian02} derived a
10 day period from their 5.3 yr optical monitoring.
The latter authors also discovered a variation range of about
3 mag in the $V$ band and 2.5 mag in both $R$ and $I$ bands during their
whole monitoring program. The optical and radio behavior of the object was
recently presented by \citet{raiteri03} based on 8 years
of optical and more than 20 years of radio observations. Four major optical
outbursts were observed at the beginning of 1995, in late 1997, at the end
of 2000, and in fall 2001. An exceptional brightening of 2.3 mag in 9 days
was detected in the R band just before a BeppoSAX pointing on Oct 30th,
2000. The radio flux
variations at different frequencies are similar, but the amplitude decreases
with increasing wavelength. Its multiwavelength variability was described
in detail in \citet{wagner96}.

The broad band spectral properties of S5 0716+714 suggest that it is
intermediate between low-frequency peaked BL Lac (LBL) and high-frequency
peaked BL Lac (HBL) \citep{ulrich97}.
The high-energy part of its spectral energy distribution is expected to
peak in the MeV energy domain. A 450 ks {\sl INTEGRAL} observation of S5
0716+714 proposed by the Landessternwarte Heidelberg group (PI: S. Wagner)
was performed in the period 2003 November 10--18.
A number of ground-based radio and optical telescopes have monitored
this source during this period, including our 60/90 Schmidt telescope.

Our monitoring program have covered the period from 2003 November 8 to 18.
and from 2003 December 30 to 2004 January 5 with a temporal resolution of
about 20 minutes. Unlike most previous investigations, which focused on
the long-term variability of this object, we concentrated on its
microvariability in short timescales based on our high temporal
resolution. The short timescales and spectral behaviors were studied.
Here we present the observational results and analysis.

This paper is organized as follows: The observation and reduction
procedures on the monitoring data are described in Sect.~2. Sect.~3
presented the results, including the light curves, the analysis on
timescales and spectral changes, and some discussion on the sine-like
light curves we observed. A summary is given in Sect.~4.

\section{OBSERVATION AND DATA REDUCTION}

Our optical monitoring program was performed on a 60/90 Schmidt telescope,
which is located at the Xinglong Station of the
National Astronomical Observatories of China (NAOC). A Ford Aerospace
$2\,048\times2\,048$ CCD camera is mounted at its main focus. The CCD has
a pixel size of 15 microns and its field of view is $58'\times58'$,
resulting in a resolution of 1.7 \arcsec/pixel. The telescope is equipped
with a 15-color intermediate-band photometric system, covering a wavelength
range from 3\,000 to 10\,000 \AA. The telescope and the photometric system
are mainly used to carry out the Beijing-Arizona-Taiwan-Connecticut (BATC)
survey and has shown their efficiency in detecting fast variabilities in
blazars \citep[e.g.,][]{peng03}.

The observations on the telescope are now highly automated. The telescope
and filters can be controlled by a single computer command with a parameter
file that specifies the telescope pointing, the filter change, the exposure
time, etc. Once the observation starts, The remaining work left for the
observers are to check the quality of the observed CCD images and
to pay some attention to the weather condition. In fact, after the night
assistant prepared the hardwares, the monitoring of S5 0716+714 was
controlled remotely by a computer at the headquarters of NAOC
in Beijing, which is about 140 km away from the telescope.

Our monitoring of S5 0716+714 was divided into two periods, one was from
2003 November 8 to 18 (6 nights, in fact, due to weather condition), the
other was from 2003 December 30 to 2004 January 5 (7 successive nights). The
first period (hereafter Period 1) covered the duration of the {\sl INTEGRAL}
observation and the second period (hereafter Period 2) was an extension of
Period 1. A filter cycle of $e$, $f$, and $k$ (their central
wavelengths are 4873, 5248, and 7528 \AA, respectively) was used in
Period 1 and a typical exposure time of 200--300 s was able to produce
an image with a good signal-to-noise ratio. Only the central $512\times512$
pixels were read out as the CCD images and the readout time was about
5.6 s. Plus the time on filter change, we achieved a temporal resolution
of about 20 min in each band. This enabled us to realize quasi-simultaneous
measurements in three BATC bands with a high temporal resolution by using
only one telescope. The size of the $512\times512$ image is
$14\farcm5\times14\farcm5$ and is enough to cover the BL Lac and eight
previously published comparison stars \citep{villata98}.

\begin{figure}
\plotone{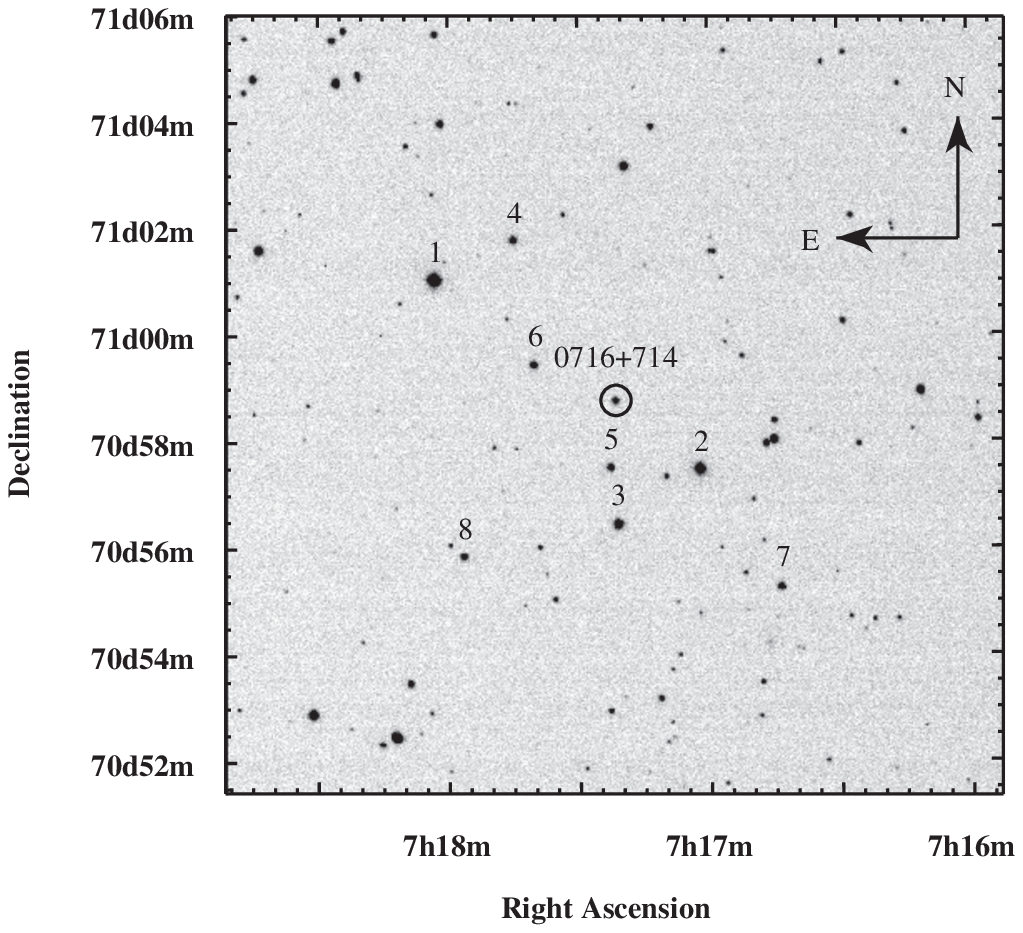}
\caption{Finding chart of S5 0716+714 and the 8 comparison stars taken
with the 60/90 Schmidt telescope and filter $i$ on JD~2\,453\,008. The
size is $14\farcm5\times14\farcm5$ (or 512$\times$512 in pixels). The 8
comparison stars are the same as in \citet{villata98}.}
\label{F1}
\end{figure}

During Period 2, we changed the $k$ filter to the more sensitive
$i$ filter (its central wavelength is 6711 \AA), with which a shorter
exposure time can produce images with the same quality as in the $k$
band. The other observational procedures and constraints were the same as
in Period 1.
The observational log and parameters are presented in Tables~2--7 with
columns being observation date and time (UT), exposure time, Julian
Date, BATC magnitude and error. The finding chart of S5 0716+714 and
the comparison stars is illustrated in Fig.~\ref{F1}.

In order to obtain in real time the light curves of the BL Lac object
we developed an automatic procedure. The procedure includes the
following steps. The CCD images were first flat-fielded, then Bertin's
Source EXtractor \citep[SEX,][]{bertin96} was run on the CCD frames and
the instrumental magnitudes and errors of S5 0716+714 and the 8 comparison
stars were extracted. The average FWHM of the stellar images is about
4.0$\arcsec$. A photometric aperture of 5 pixel (8.5$\arcsec$)
diameter was adopted during the extraction. The BATC $e$, $f$, $i$, and
$k$ magnitudes of the 8 comparison stars were obtained by observing them
and a BATC standard star HD~19445 in a same night and are listed in
Table~1. Then, by comparing the instrumental
magnitudes of the 8 comparison stars with their BATC magnitudes,
the instrumental magnitudes of S5 0716+714 were calibrated into the BATC
$e$, $f$, $i$, and $k$ magnitudes and the light curves in
4 BATC bands were obtained.

\section{RESULTS}
\subsection{Light Curves}
The light curves are displayed in Figs.~2 and 3 for Periods 1 and 2,
respectively. The large panels
show light curves of the BL Lac and the small ones present those
of the differential magnitudes (average set to 0) between the 5th
comparison star and the average of all 8. In order to show the variation
clearly, we only plot the duration when the BL Lac was observed and
exclude the daytime periods when no observations could be made by us.

\begin{figure}
\epsscale{.55}
\plotone{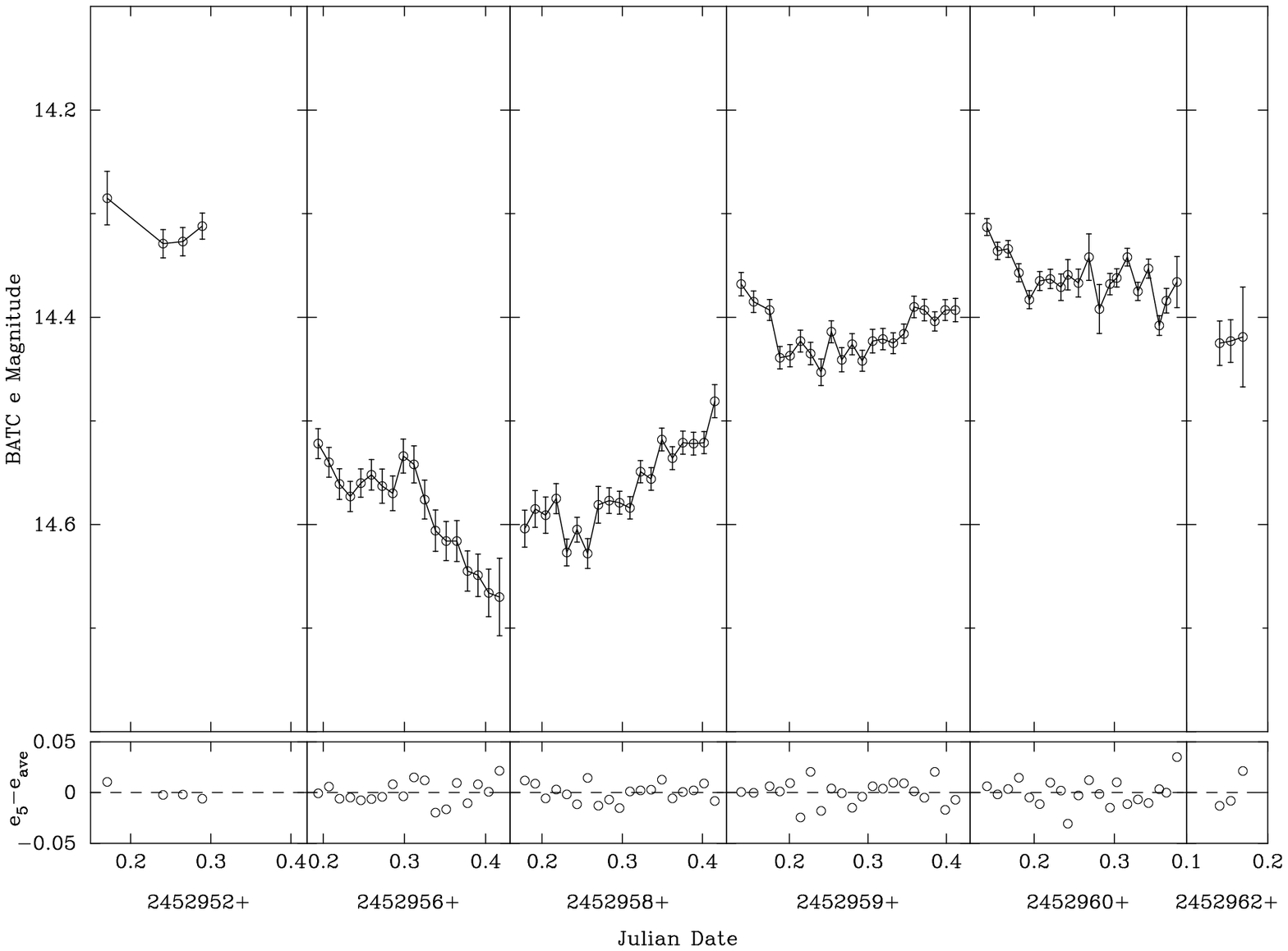}
\plotone{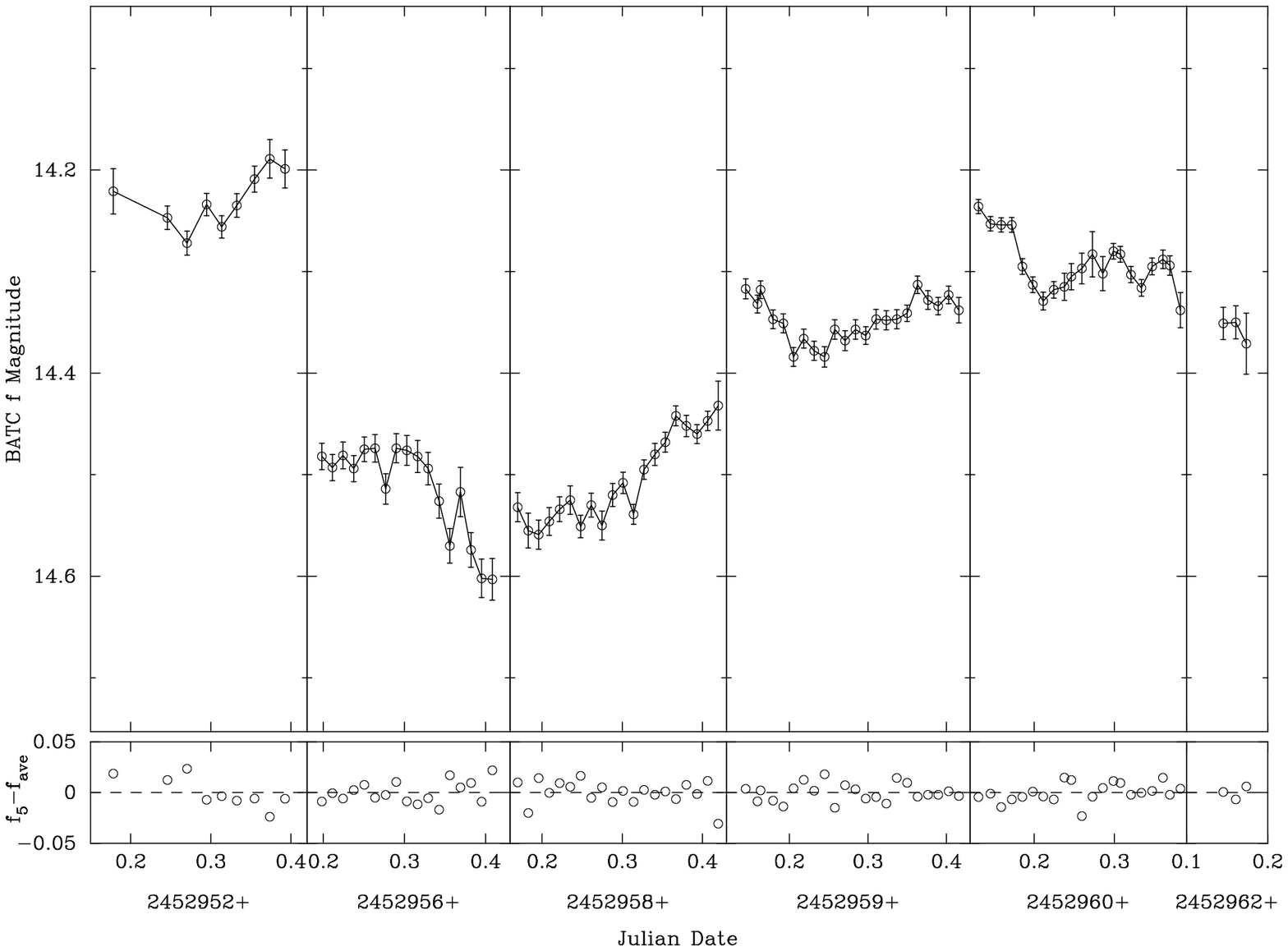}
\plotone{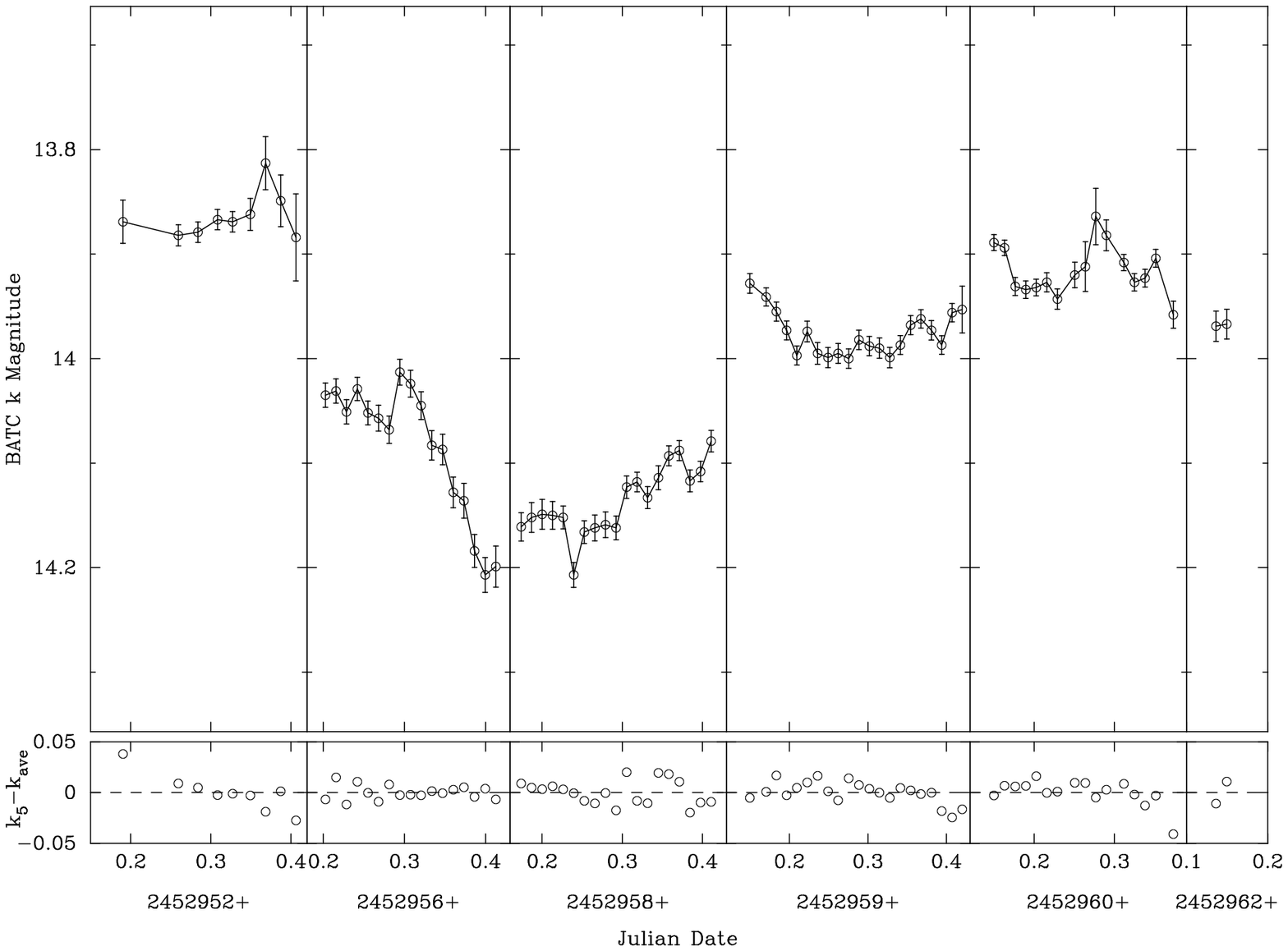}
\caption{Light curves of S5 0716+714 in the BATC $e$, $f$, and $k$ bands
in Periods 1. Large panels are those of the BL
Lac object and small ones are of the differential magnitudes between
the 5th comparison star and the average of all 8.}
\label{F2}
\end{figure}

All light curves show intranight fluctuations superposed on longer
timescale variations. In Period 1 (see Fig.~\ref{F2}), the
variation is characterized by fast oscillations with
small amplitudes. In the first night or on JD~2\,452\,952, the BL Lac
was in a relatively `high' state. Then its brightness dropped in
the following days and reached a minimum on JD~2\,452\,956.
In the following two days, the BL Lac got brighter,
reaching another `high' state around the beginning of JD~2\,452\,959.
The total magnitude change was about 0.4 mag in Period 1 and the
magnitude changes in individual nights were mostly about 0.1 mag except
that on JD~2\,452\,956. The object appeared in a relatively quiescent state.

The observational accuracy in Period 2 is higher than in Period 1 due to
the better weather conditions in Period 2. The light curves are
characterized by continuous (except on JD~2\,453\,006) increase in
brightness. Two outbursts were observed on JD~2\,453\,005 and
JD~2\,453\,009 with rapid brightening of more than 0.3 mag within 0.4
days. The most sharp increase in brightness occurred on JD~2\,453\,005.
The $i$ magnitude changed from 13.896 on JD~2453005.328 to 13.716 on
JD~2453005.400 (see Table~7), resulting in a rising rate of 0.1 mag per hour.
The total magnitude change is about 0.8 mag in Period 2, which is a
factor of two larger than that in Period 1, and the object appeared in
an active or flaring state.

\begin{figure}
\epsscale{.55}
\plotone{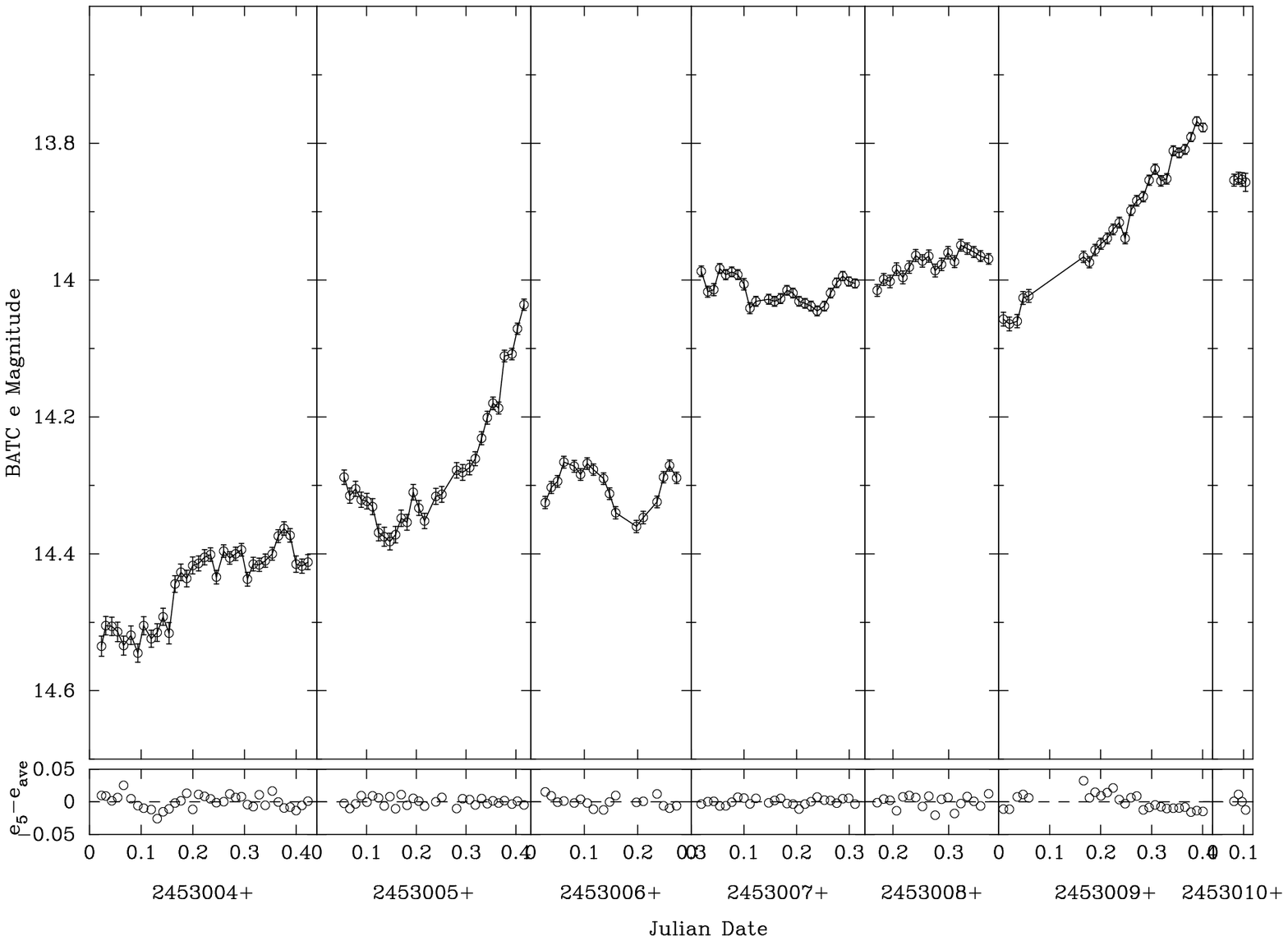}
\plotone{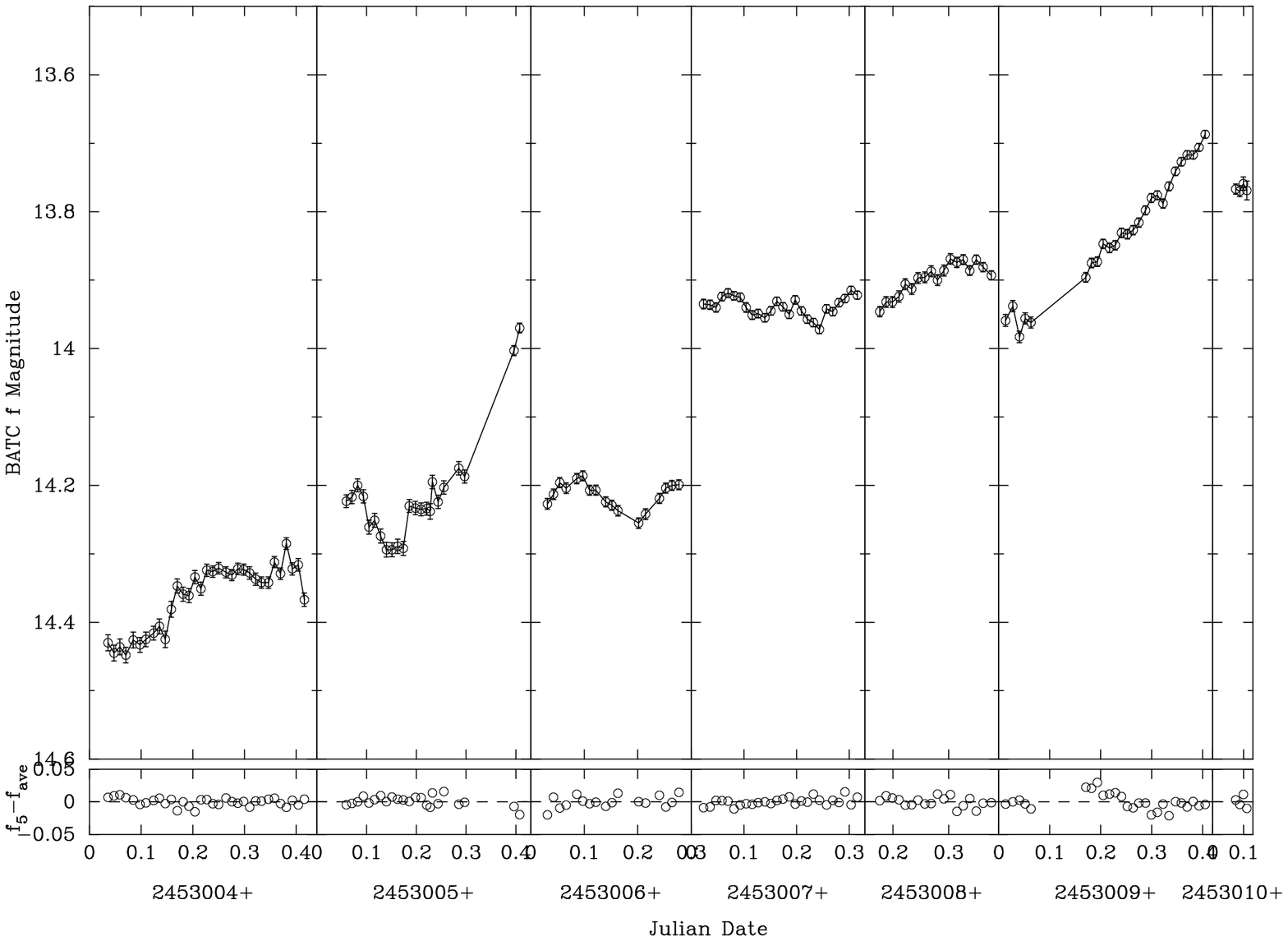}
\plotone{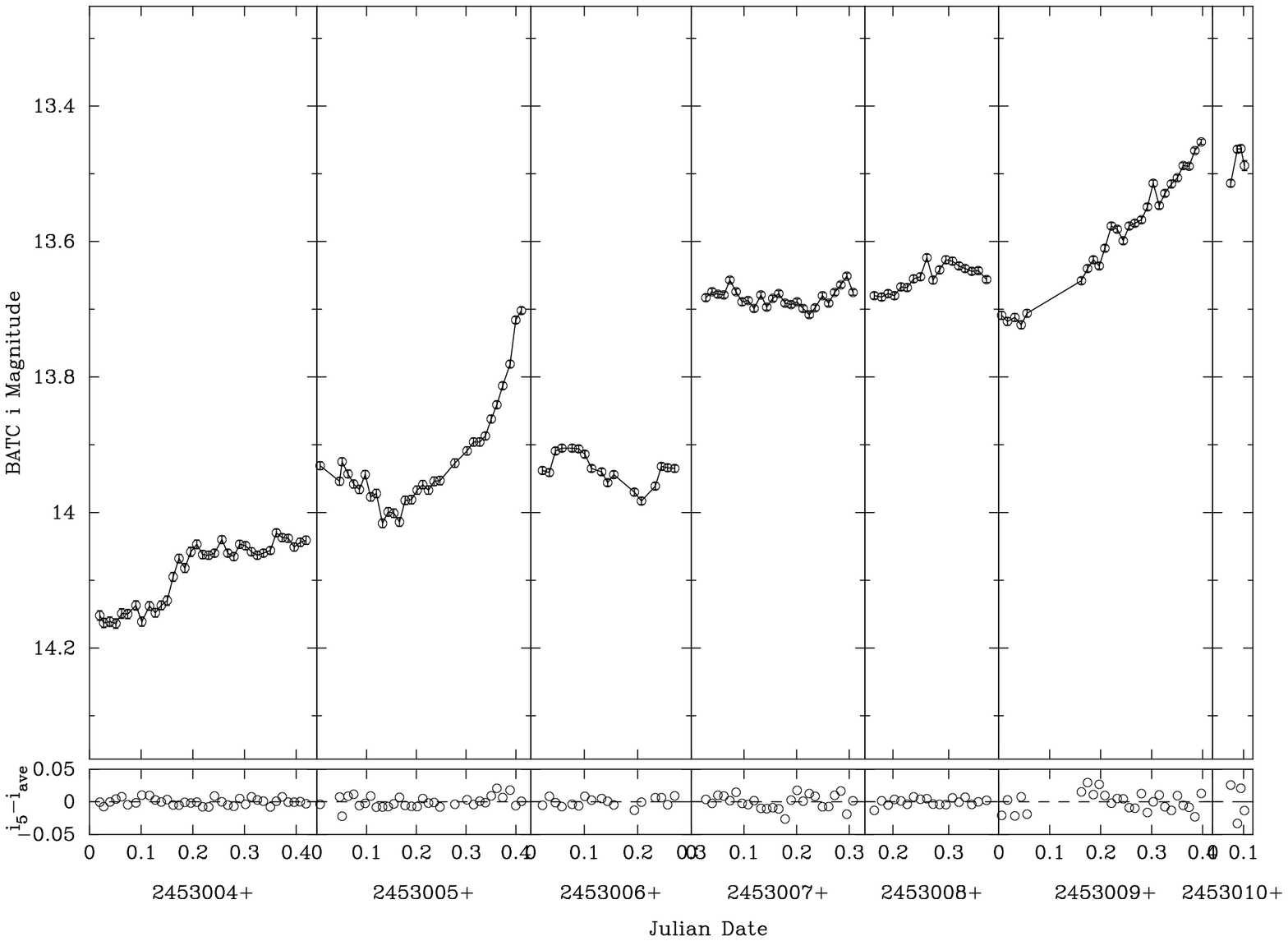}
\caption{Light curves of S5 0716+714 in the BATC $e$, $f$, and $i$ bands
in Period 2. Large panels are those of the BL
Lac object and small ones are of the differential magnitudes between
the 5th comparison star and the average of all 8.}
\label{F3}
\end{figure}

The most unusual variation in Period 2 was observed on JD~2\,453\,006:
all three light curves look very close to sine curves (see Fig.~\ref{F3})
with an amplitude of about 0.1 mag and a period of about 0.21 day (5 hr).
This kind of variation is of particular interest and will be discussed
in Sect.~3.4.

In order to establish whether there is a time lag between the variations
in different wavebands, we have calculated the $z$-transformed
discrete correlation function \citep[ZDCF,][]{alex97} for
Periods 1 and 2 and for several individual nights. No significant time
lag has been identified except that a couple of nights show time lags from
a few to less than 20 minutes between different wavebands. A time lag between
variations in different wavebands will lead to an oscillating color
index with respect to brightness, rather than a bluer-when-brighter
trend reported in Sect.~3.3.

In both periods, the light curves in different bands are consistent with
one another. The rms's of the differential magnitudes between the 5th
comparison star and the average of all 8 are 0.011, 0.010, 0.012,
0.009, 0.008, and 0.010 mag in the six small panels in Figs.~2 and 3.
These results demonstrate the accuracy of our magnitude measurements.

\subsection{Timescales of Variability}
Structure function \citep{simon85} can be
used to search for the typical timescales and periodicities of the
variability. A characteristic timescale in a light curve, defined as
the time interval between a maximum and an adjacent minimum or vice versa,
is indicated by a maximum of the structure function, whereas a periodicity
in the light curve causes a minimum of the structure function \citep{heidt96}.

\begin{figure}
\epsscale{.90}
\plottwo{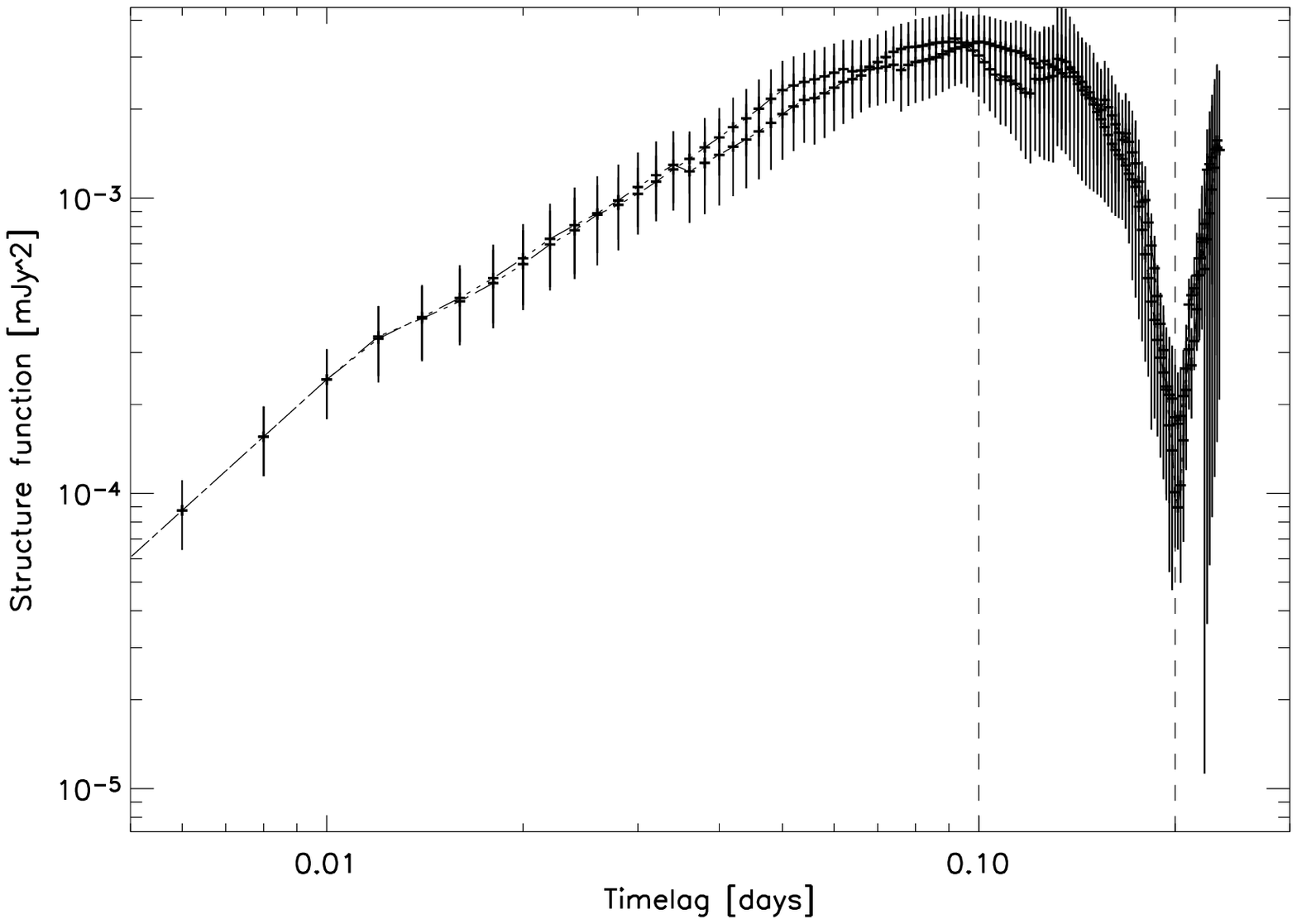}{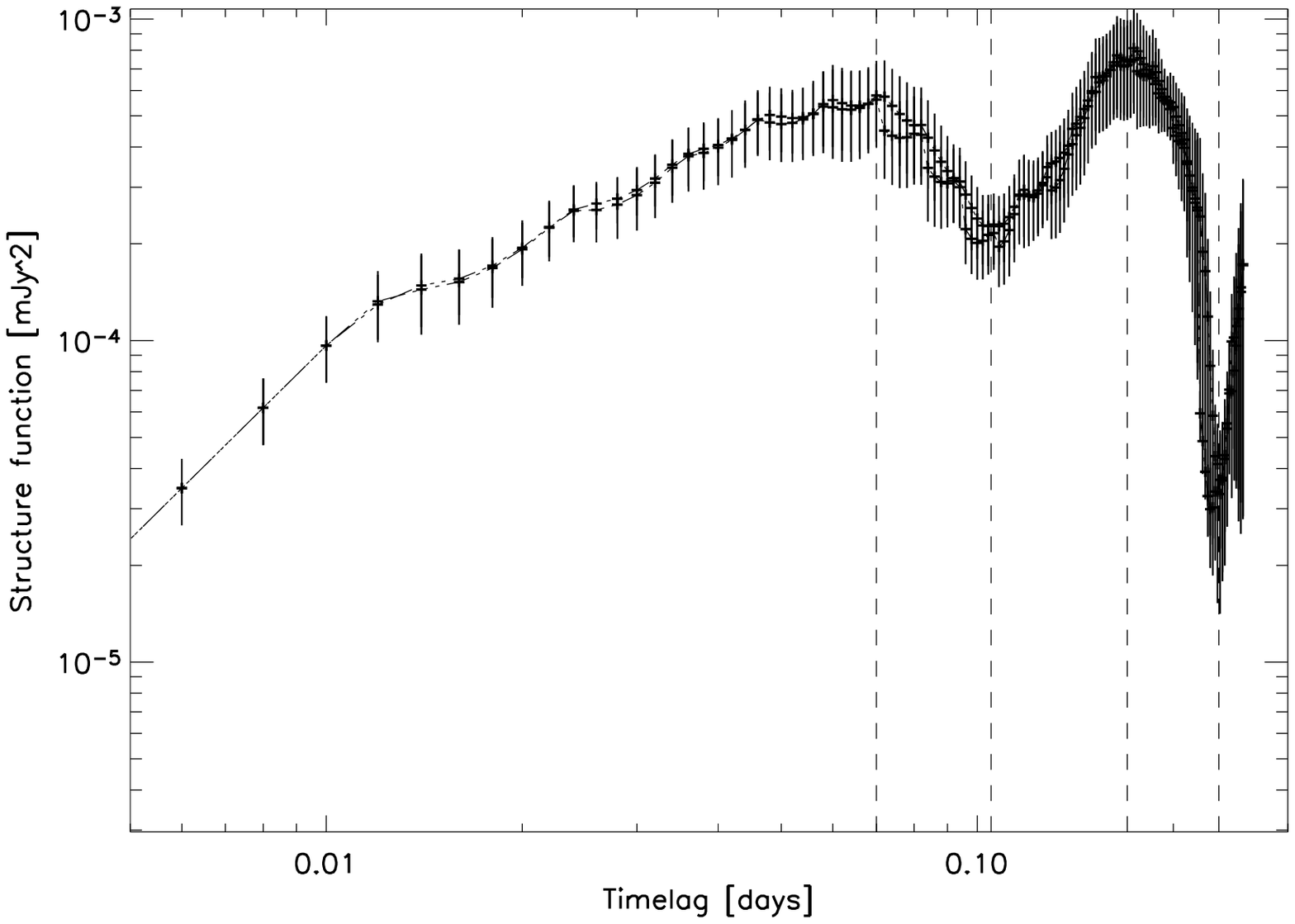}
\caption{Structure function of the S5 0716+714 on JD~2\,453\,006 (the
left panel) and JD~2\,453\,007 (the right panel). The dashed lines
indicate timescales at maxima and periods at minima of structure functions.}
\label{F4}
\end{figure}

For S5 0716+714, structure function analysis was performed on the light
curves of each individual night. Short timescales of a few hours were
derived but the results are different from night to night. For example,
the structure function analysis (see the left panel in Fig.~\ref{F4})
identified a timescale of 0.11 day (2.5 hr) and a period of 0.21 day (5 hr)
for JD~2\,453\,006, which are consistent with the period clearly visible
on the sine-like light curves in that night. Another example is that the
same analysis on light curves of JD~2\,453\,007 revealed timescales of
0.07 and 0.17 day (1.7 and 4.1 hr) and periods of 0.11 and 0.23 day (2.6
and 5.5 hr) (see the right panel in Fig.~\ref{F4}). All timescales
and periods are labeled by dashed lines in Fig.~\ref{F4}. There are also
a common timescale of about 20 min appeared in all structure functions.
But this timescale is identical to the temporal resolution of our monitoring
and can not be associated with the intrinsic variability.

IDV has been frequently reported at radio and optical wavelengths in
BL Lac S5 0716+714, our observations at optical bands
re-confirmed such IDV phenomena in this source. Comparing to the much
longer optical timescales of 4--10 day derived by other
authors \citep[e.g.,][]{heidt96,qian02},
our dense monitoring enabled us to derive much shorter timescales
for this object, which may constrain the physical processes
that result in the fast microvariability of S5 0716+714 (see discussion
in Sect.~4).

\subsection{Spectral Behavior}
The optical spectral change with brightness has been investigated
for S5 0716+714 \citep[e.g.,][]{ghisellini97,raiteri03,villata00,
villata04} and for other BL Lac objects \citep[e.g.,][]{carini92,
romero00,speziali98,villata02}. Most authors have reported a
bluer-when-brighter chromatism when the objects are in fast flares
and an ``achromatic" trend for their long-term variability. However,
\citet{raiteri03} also noted for S5 0716+714 that, on short timescales,
``different behaviours have been found: sometimes a bluer-when-brighter
trend is recognizable, while in some other cases the opposite is true;
there are also cases where magnitude variations do not imply spectral
changes". They suggested to perform a very dense monitoring with high
precision data to distinguish trends in the short-term
spectral behaviour of this source.

\begin{figure}
\epsscale{.90}
\plottwo{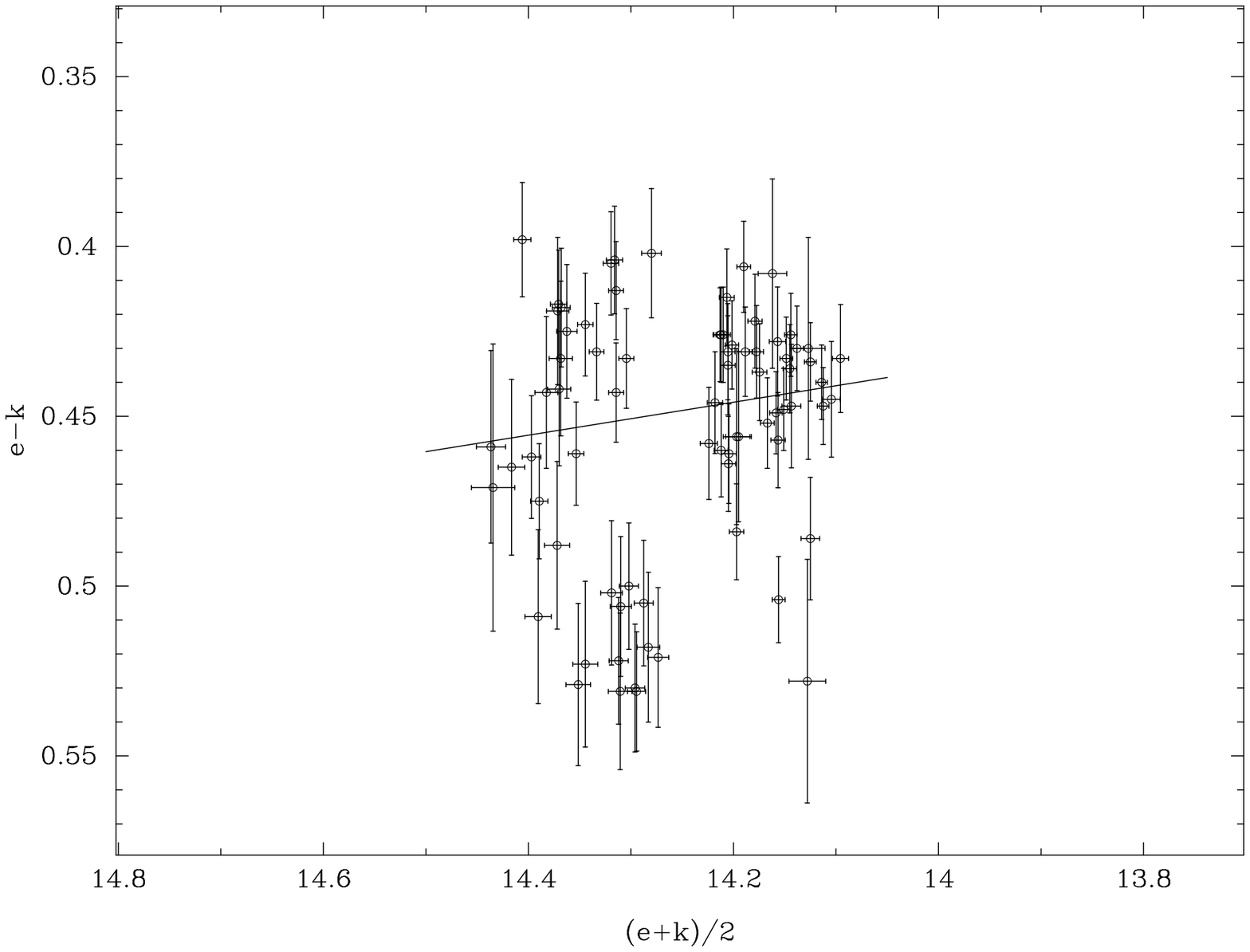}{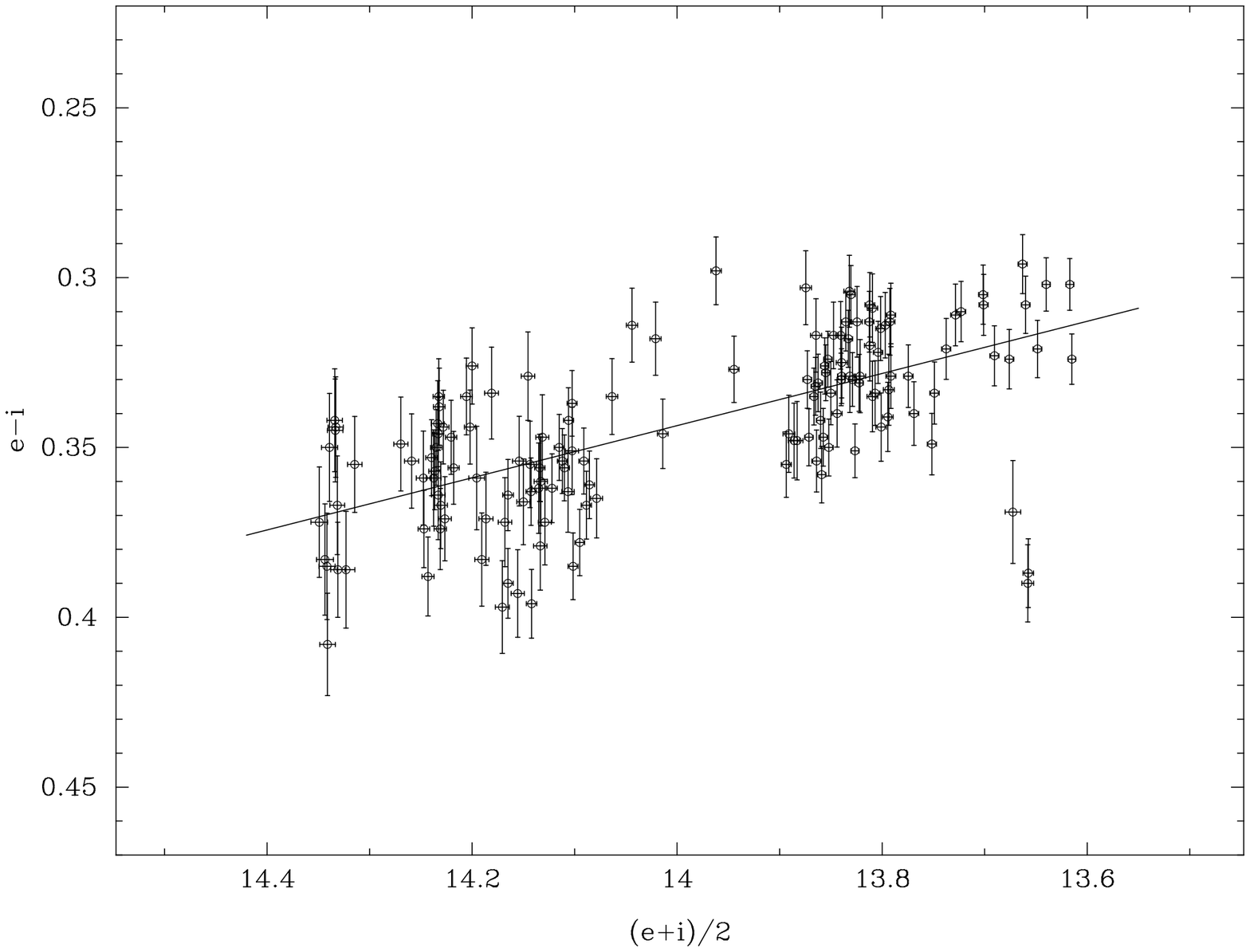}
\caption{Color index vs brightness in two periods. The lines are
best fits to the points. The bluer-when-brighter chromatism is clear
in Period 2 (right panel) but absent in Period 1 (left panel). The
two panels are set to have the same scales on both coordinates for a
comparison.}
\label{F5}
\end{figure}

Our monitoring of S5 0716+714 with high temporal resolution enables
us to study its spectral behavior with a high confidence level.
Following most authors mentioned above, we use color index to denote
the spectral shape. The color index and brightness are taken
as $e-k$ and $\frac{e+k}{2}$ for Period 1 and $e-i$ and $\frac{e+i}{2}$
for Period 2. The changes of the color index to the brightness in two
periods are shown in Fig.~\ref{F5}. The solid lines are best
fits to the points and have taken the errors in both coordinates into
consideration \citep{press92}.

In Period 1 or when the object was in a relatively quiescent
state, The distribution in color-brightness diagram
is quite dispersed and there was not an overall color change. However, in
Period 2 or when the object was in a flaring state, a clear
bluer-when-brighter chromatism was found. The linear fit has a slope
of 0.077. The Pearson correlation coefficient is 0.636 and the
significance level is $5.014\times10^{-18}$, suggesting a strong
correlation between color index and brightness. This is in agreement
with the results obtained by most authors mentioned above.
\citet{ghisellini97} have deduced from their monitoring that two processes
may be operating in this source: the first one would cause the achromatic
long-term flux variations, while the second would be responsible for the
short-term fast variations. Our results are somewhat different but still
consistent with theirs: during the quiescent or low state, the variation
may be dominated by the long-term component and shows no clear spectral
change; while in the active or flaring state, the variation is dominated by
the short-term component and has a bluer-when-brighter chromatism. 

That the spectra of S5 0716+714 change with its brightness
have been observed in other wavelengths. For instances, \citet{raiteri03}
reported a flatter-when-brighter trend in radio waveband and \citet
{cappi94} discovered a steeper-when-fainter phenomenon in soft X-ray.
In fact, the spectrum changing with the flux in multi-wavelengths is a
common feature
in blazars \citep[e.g.,][]{aller85,urry86,sembay93,kniffen93,mukherjee96}.
This universal spectral behaviour is nontrivial. It suggests a close
relationship among the mechanisms that are responsible for the emission
and variation in different wavebands. The analysis of spectral
changes of blazars can put some strong constraints on the physical
processes that are responsible for their variations (see discussion
in Sect.~3.4).

\subsection{The Sine Light Curves}
The perfect sine light curves observed on JD~2\,453\,006 is of particular
interest because very few of this kind of light curve has
been reported before. They mimic a periodic variation but there is only
one complete period (it's a great pity that our weather got bad just at
the end of this period). \citet{webb90} has detected a sinusoidal component
in the variations of 3C~120, but the period was much longer ($\sim13$ yr).
For S5 0716+714, quasi-periodic oscillations have been detected
by \citet{quir91}. Their light curves, however, are
very different: they are sawtooth-like with sharp turnoffs while our
light curves are sine-like with smooth turns. Then we come to the
question: what mechanism can produce such sine light curves?

Some mechanisms have been proposed to explain the IDV phenomena of blazars.
They can be largely classified into two types, the extrinsic and the
intrinsic, as reviewed by \citet{wagner95}. The extrinsic mechanisms
include the inter-stellar scintillation (ISS) and gravitational
microlensing. The ISS is highly frequency-dependent
and only operates at low radio frequencies. The IDV in the mm regime and
the fast variability in the optical regime, as observed by us, cannot
be caused by the ISS mechanism. Meanwhile, the microlensing is an
achromatic process and will result in symmetric light curves. However,
color or spectral changes have been frequently observed
from radio to X-ray wavebands for the variability of S5 0716+714, as
mentioned in last section. We have also detected a clear bluer-when-brighter
chromatism. In addition, the light curves at all wavelengths, including
our optical ones, are generally asymmetric. Therefore, the fast
variation of S5 0716+714 is unlikely due to microlensing. 
In fact, the close correlation between the optical and radio bands observed
in S5 0716+714 \citep[e.g.,][]{quir91,wagner96} provides a strong
evidence against the extrinsic origin of its variability.

The intrinsic interpretations include mainly the accretion disk
instabilities and shock-in-jet model. The accretion disk model is able
to explain some of the phenomena seen in the optical to X-ray range but
cannot explain radio IDV \citep[e.g.,][]{wagner95}.
The most frequently cited model is the shock-in-jet model which has been
widely used to explain the variability of blazars and quasars
\citep[e.g.,][and references therein]{guetta04,jia98,qian91,romero00,
wagner95}. The main idea of the model is
that shocks propagate down the relativistic jet whose plasma is
hydromagnetically turbulent. At sites where the shocks encounter particles
or magnetic field overdensities, the optical synchrotron emission is
enhanced. The amplitude and timescale of the
resulting variation depend on the power spectrum of the turbulence
and the shock thickness. This kind of shock-in-jet model will naturally
lead to the prediction of a bluer-when-brighter phenomenon
\citep{marscher98}, as observed in our case.

The shock-in-jet model still suffers from a number of problems in
explaining IDV, such as the close correlation between the radio and
optical variations. Therefore, the geometrical effects are sometimes
invoked to account for some observational facts that can not be
interpreted satisfactorily by the shock-in-jet model. Geometrical
modulation in the context of shock-in-jet models are detailed
by \citet{camen92}. They argued that knots
of enhanced particle density are injected at a finite jet radius. In
knots moving relativistically on helical trajectories, the direction
of forward beaming varies with time. For an observer close to the jet
axis, the sweeping of the beam will introduce flares due to the light
house effect. This will lead to quasi-periodic variations of a few
oscillations and the variations are basically achromatic.

\begin{figure}
\epsscale{.90}
\plotone{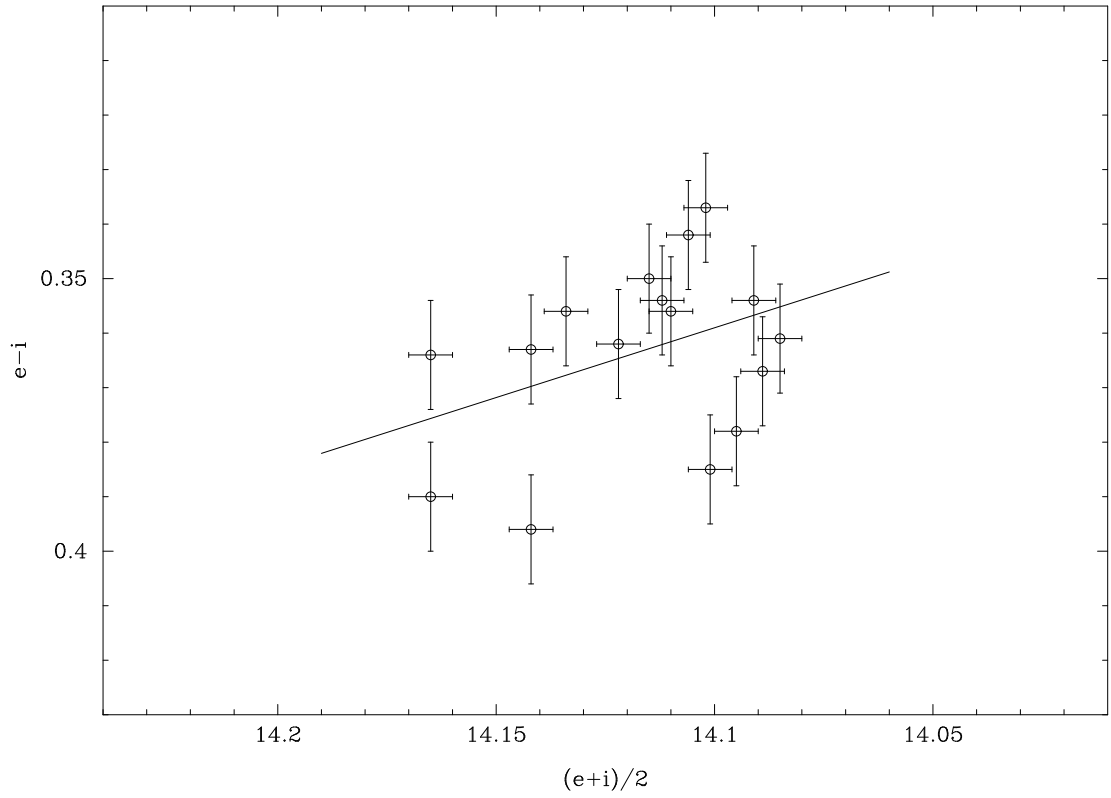}
\caption{Color index vs brightness on JD~2\,453\,006. The solid line
only indicates a very weak correlation between the color index and
brightness.}
\label{F6}
\end{figure}

It is tempting to examine the color change on only JD~2\,453\,006 since
perfect sine-like light curves were observed in that night. Fig.~\ref{F6}
illustrates the color index vs brightness relation. The linear fit gives
a slope of 0.256 which is very different from the overall slope of
Period 2. The correlation coefficient is 0.361 and the significance
level is 0.170, which means a poor fit or no clear correlation between
the color index and brightness. That is to say, the brightness changed
nearly achromatically on JD~2\,453\,006. The only two known processes
that can cause achromatic variability are the microlensing and light house
effect. Although the microlensing has been ruled out as the dominant
mechanism of the variability of S5 0716+714, it may still have some
contribution. The symmetry in the sine light curves may indicate a
microlensing event, but the concave shape of the second halves of the light
curves can not be explained in terms of microlensing. In addition, this
very short timescale would require a transverse speed of $v_{trans}\sim
c$ when microlensing. Therefore, the most probable mechanism responsible
for the sine light curves is the light house effect. It
may produce a periodic variation according to \citet{camen92} and the
variation is achromatic. In other words, the variation observed
on JD~2\,453\,006 is likely due to geometrical effects.

It is unclear whether all fast IDV, especially the
quasi-periodic ones, such as those observed by
\citet{quir91}, can also be explained in terms of geometrical effects
within the context of shock-in-jet model. If the answer is yes,
the actual timescales of the intrinsic flux changes will be longer
by a few factors and the deduced extremely high brightness
temperature will be reduced by one or more orders of magnitude. This will
help to resolve the large difference between the high brightness
temperature ($\sim10^{17}$ K) and the Compton limit ($<10^{12}$ K),
though not fully resolved.

\section{SUMMARY}
During the periods of 2003 November 8--18 and from 2003 December 30 to
2004 January 5, we have carried out optical monitoring of the BL Lac
object S5 0716+714 with a high temporal resolution. Intraday and
intranight variations were observed with an overall magnitude change
of about 0.9 mag during the whole campaign. Two outbursts were recorded
on JD~2\,453\,005 and JD~2\,453\,009. Short timescales of a few
hours were derived from the light curves of each individual night but are
different from night to night. A bluer-when-brighter chromatism was present
when the object was in an active or flaring state but was absent when
it was in a relatively low or quiescent state.

Our observations have suggested that the fast microvariability in
S5 0716+714 is basically consistent with the shock-in-jet model. The
analysis has also indicated that the geometrical effects can sometimes
play an important role in the variability of blazars. Up to now, all
theoretical models that have been proposed to explain the variability
of blazars have their own individual difficulties \citep[see, e.g.,][for
a discussion]{wagner95}. In order to better understand the variability
of blazars and to strictly constrain the theoretical models, simultaneous
multifrequency campaigns with high temporal resolutions should be the
direction of future efforts.

\begin{acknowledgements}
The authors thank the anonymous referee for valuable comments and suggestions
that helped to improve this paper very much. This work has been supported
by the Chinese National Key Basic Research Science Foundation (NKBRSF
TG199075402) and in part by the Chinese National Science Foundation, No.
10303003 and No. 10473012. Peng acknowledges grant NKBRSF2003CB716703 and
NSF grant No. 10173015.
\end{acknowledgements}

\clearpage

\clearpage
\clearpage
\clearpage
\clearpage
\clearpage
\clearpage

\oddsidemargin=-1cm
\tabletypesize{\scriptsize}

\begin{deluxetable}{lcccc}
\tablewidth{0pt}
\tablecaption{BATC Magnitudes of 8 Comparison Stars.\label{T1}}
\tablehead{\colhead{Star} & \colhead{$e$} & \colhead{$f$} & \colhead{$i$} & \colhead{$k$}}
\startdata
 1 & 11.207 & 11.174 & 11.075 & 11.007 \\
 2 & 11.656 & 11.615 & 11.508 & 11.490 \\
 3 & 12.663 & 12.564 & 12.428 & 12.361 \\
 4 & 13.366 & 13.296 & 13.245 & 13.211 \\
 5 & 13.760 & 13.661 & 13.511 & 13.449 \\
 6 & 13.839 & 13.747 & 13.590 & 13.564 \\
 7 & 14.049 & 13.962 & 13.628 & 13.645 \\
 8 & 14.335 & 14.284 & 14.113 & 14.106 \\
\enddata
\end{deluxetable}
\clearpage
\begin{deluxetable}{cccccc}
\tablecaption{Observational log and results in the BATC $e$ band in Period 1.\label{T2}}
\tablewidth{0pt}
\tablehead{
\colhead{Obs. Date\tablenotemark{a}} & \colhead{Obs. Time\tablenotemark{a}} & \colhead{Exp. Time} & \colhead{JD} & \colhead{$e$} & \colhead{$e_{\rm err}$} \\
\colhead{(yyyy mm dd)} & \colhead{(hh:mm:ss)} &  \colhead{(s)} & & \colhead{(mag)} & \colhead{(mag)}
}
\startdata
2003 11 08 & 16:04:57 & 100 & 2452952.171 & 14.285 & 0.026 \\
2003 11 08 & 17:43:46 & 300 & 2452952.241 & 14.329 & 0.014 \\
2003 11 08 & 18:19:14 & 300 & 2452952.265 & 14.327 & 0.014 \\
2003 11 08 & 18:54:11 & 300 & 2452952.289 & 14.312 & 0.013 \\
2003 11 12 & 16:36:29 & 300 & 2452956.194 & 14.522 & 0.014 \\
2003 11 12 & 16:55:35 & 300 & 2452956.207 & 14.540 & 0.014 \\
2003 11 12 & 17:14:19 & 300 & 2452956.220 & 14.561 & 0.015 \\
2003 11 12 & 17:33:15 & 300 & 2452956.233 & 14.573 & 0.015 \\
2003 11 12 & 17:52:09 & 300 & 2452956.246 & 14.560 & 0.014 \\
2003 11 12 & 18:10:59 & 300 & 2452956.259 & 14.552 & 0.015 \\
2003 11 12 & 18:29:50 & 300 & 2452956.273 & 14.563 & 0.016 \\
2003 11 12 & 18:48:47 & 300 & 2452956.286 & 14.570 & 0.017 \\
2003 11 12 & 19:07:32 & 300 & 2452956.299 & 14.534 & 0.016 \\
2003 11 12 & 19:26:31 & 300 & 2452956.312 & 14.542 & 0.018 \\
2003 11 12 & 19:45:18 & 300 & 2452956.325 & 14.576 & 0.019 \\
2003 11 12 & 20:04:36 & 300 & 2452956.338 & 14.606 & 0.020 \\
2003 11 12 & 20:23:30 & 300 & 2452956.351 & 14.616 & 0.019 \\
2003 11 12 & 20:42:24 & 300 & 2452956.364 & 14.616 & 0.020 \\
2003 11 12 & 21:01:19 & 300 & 2452956.378 & 14.645 & 0.019 \\
2003 11 12 & 21:20:02 & 300 & 2452956.391 & 14.649 & 0.021 \\
2003 11 12 & 21:38:58 & 300 & 2452956.404 & 14.666 & 0.023 \\
2003 11 12 & 21:58:10 & 300 & 2452956.417 & 14.670 & 0.037 \\
2003 11 14 & 16:14:20 & 300 & 2452958.178 & 14.604 & 0.018 \\
2003 11 14 & 16:33:04 & 300 & 2452958.191 & 14.585 & 0.018 \\
2003 11 14 & 16:51:59 & 300 & 2452958.205 & 14.591 & 0.018 \\
2003 11 14 & 17:10:48 & 300 & 2452958.218 & 14.575 & 0.014 \\
2003 11 14 & 17:29:48 & 300 & 2452958.231 & 14.627 & 0.013 \\
2003 11 14 & 17:48:29 & 300 & 2452958.244 & 14.605 & 0.012 \\
2003 11 14 & 18:07:29 & 300 & 2452958.257 & 14.628 & 0.014 \\
2003 11 14 & 18:26:39 & 300 & 2452958.270 & 14.581 & 0.018 \\
2003 11 14 & 18:45:54 & 300 & 2452958.284 & 14.577 & 0.012 \\
2003 11 14 & 19:04:43 & 300 & 2452958.297 & 14.579 & 0.011 \\
2003 11 14 & 19:23:29 & 300 & 2452958.310 & 14.584 & 0.011 \\
2003 11 14 & 19:42:20 & 300 & 2452958.323 & 14.549 & 0.011 \\
2003 11 14 & 20:01:27 & 300 & 2452958.336 & 14.556 & 0.011 \\
2003 11 14 & 20:20:35 & 300 & 2452958.349 & 14.518 & 0.011 \\
2003 11 14 & 20:39:35 & 300 & 2452958.363 & 14.536 & 0.011 \\
2003 11 14 & 20:58:20 & 300 & 2452958.376 & 14.521 & 0.011 \\
2003 11 14 & 21:17:37 & 300 & 2452958.389 & 14.522 & 0.011 \\
2003 11 14 & 21:36:34 & 300 & 2452958.402 & 14.521 & 0.011 \\
2003 11 14 & 21:55:39 & 300 & 2452958.415 & 14.481 & 0.016 \\
2003 11 15 & 15:17:15 & 300 & 2452959.139 & 14.368 & 0.011 \\
2003 11 15 & 15:39:57 & 300 & 2452959.154 & 14.385 & 0.010 \\
2003 11 15 & 16:09:28 & 300 & 2452959.175 & 14.393 & 0.010 \\
2003 11 15 & 16:28:10 & 300 & 2452959.188 & 14.439 & 0.011 \\
2003 11 15 & 16:46:52 & 300 & 2452959.201 & 14.437 & 0.011 \\
2003 11 15 & 17:05:52 & 300 & 2452959.214 & 14.423 & 0.011 \\
2003 11 15 & 17:24:35 & 300 & 2452959.227 & 14.435 & 0.011 \\
2003 11 15 & 17:43:40 & 300 & 2452959.240 & 14.453 & 0.013 \\
2003 11 15 & 18:02:34 & 300 & 2452959.254 & 14.414 & 0.011 \\
2003 11 15 & 18:21:30 & 300 & 2452959.267 & 14.441 & 0.012 \\
2003 11 15 & 18:40:35 & 300 & 2452959.280 & 14.426 & 0.010 \\
2003 11 15 & 18:59:20 & 300 & 2452959.293 & 14.442 & 0.010 \\
2003 11 15 & 19:18:12 & 300 & 2452959.306 & 14.423 & 0.011 \\
2003 11 15 & 19:36:58 & 300 & 2452959.319 & 14.421 & 0.010 \\
2003 11 15 & 19:56:04 & 300 & 2452959.332 & 14.425 & 0.010 \\
2003 11 15 & 20:15:17 & 300 & 2452959.346 & 14.416 & 0.009 \\
2003 11 15 & 20:34:12 & 300 & 2452959.359 & 14.390 & 0.010 \\
2003 11 15 & 20:53:04 & 300 & 2452959.372 & 14.393 & 0.010 \\
2003 11 15 & 21:12:13 & 300 & 2452959.385 & 14.404 & 0.009 \\
2003 11 15 & 21:31:09 & 300 & 2452959.398 & 14.393 & 0.010 \\
2003 11 15 & 21:50:01 & 300 & 2452959.412 & 14.393 & 0.011 \\
2003 11 16 & 15:20:41 & 300 & 2452960.141 & 14.313 & 0.008 \\
2003 11 16 & 15:39:53 & 300 & 2452960.154 & 14.336 & 0.008 \\
2003 11 16 & 15:58:38 & 300 & 2452960.168 & 14.334 & 0.008 \\
2003 11 16 & 16:18:00 & 300 & 2452960.181 & 14.357 & 0.009 \\
2003 11 16 & 16:36:50 & 300 & 2452960.194 & 14.383 & 0.009 \\
2003 11 16 & 16:55:34 & 300 & 2452960.207 & 14.365 & 0.009 \\
2003 11 16 & 17:14:35 & 300 & 2452960.220 & 14.363 & 0.009 \\
2003 11 16 & 17:33:32 & 300 & 2452960.233 & 14.371 & 0.013 \\
2003 11 16 & 17:46:00 & 300 & 2452960.242 & 14.359 & 0.015 \\
2003 11 16 & 18:04:47 & 300 & 2452960.255 & 14.367 & 0.014 \\
2003 11 16 & 18:23:48 & 300 & 2452960.268 & 14.342 & 0.022 \\
2003 11 16 & 18:42:33 & 300 & 2452960.281 & 14.392 & 0.024 \\
2003 11 16 & 19:01:31 & 300 & 2452960.294 & 14.368 & 0.010 \\
2003 11 16 & 19:14:01 & 300 & 2452960.303 & 14.362 & 0.009 \\
2003 11 16 & 19:32:45 & 300 & 2452960.316 & 14.342 & 0.009 \\
2003 11 16 & 19:51:44 & 300 & 2452960.329 & 14.375 & 0.009 \\
2003 11 16 & 20:11:02 & 300 & 2452960.343 & 14.353 & 0.009 \\
2003 11 16 & 20:30:16 & 300 & 2452960.356 & 14.408 & 0.009 \\
2003 11 16 & 20:43:00 & 300 & 2452960.365 & 14.384 & 0.012 \\
2003 11 16 & 21:02:02 & 300 & 2452960.378 & 14.366 & 0.025 \\
2003 11 18 & 15:20:20 & 300 & 2452962.141 & 14.425 & 0.022 \\
2003 11 18 & 15:39:24 & 360 & 2452962.154 & 14.423 & 0.021 \\
2003 11 18 & 16:01:25 & 300 & 2452962.169 & 14.419 & 0.048 \\
\enddata
\tablenotetext{a}{The obs. date and time are of universal time. They are the
same for Tables 3-7.}
\end{deluxetable}
\clearpage
\begin{deluxetable}{cccccc}
\tablecaption{Observational log and results in the BATC $f$ band in Period 1.\label{T3}}
\tablewidth{0pt}
\tablehead{
\colhead{Obs. Date} & \colhead{Obs. Time} & \colhead{Exp. Time} & \colhead{JD} & \colhead{$f$} & \colhead{$f_{\rm err}$} \\
\colhead{(yyyy mm dd)} & \colhead{(hh:mm:ss)} &  \colhead{(s)} & & \colhead{(mag)} & \colhead{(mag)}
}
\startdata
2003 11 08 & 16:16:03 & 100 & 2452952.178 & 14.221 & 0.022 \\
2003 11 08 & 17:51:25 & 300 & 2452952.246 & 14.247 & 0.012 \\
2003 11 08 & 18:26:52 & 300 & 2452952.270 & 14.272 & 0.012 \\
2003 11 08 & 19:01:50 & 300 & 2452952.295 & 14.234 & 0.011 \\
2003 11 08 & 19:29:07 & 300 & 2452952.314 & 14.256 & 0.011 \\
2003 11 08 & 19:56:22 & 300 & 2452952.333 & 14.235 & 0.012 \\
2003 11 08 & 20:28:06 & 300 & 2452952.355 & 14.209 & 0.013 \\
2003 11 08 & 20:55:32 & 300 & 2452952.374 & 14.189 & 0.019 \\
2003 11 08 & 21:22:45 & 300 & 2452952.393 & 14.199 & 0.019 \\
2003 11 12 & 16:43:00 & 300 & 2452956.198 & 14.482 & 0.013 \\
2003 11 12 & 17:01:49 & 300 & 2452956.211 & 14.493 & 0.013 \\
2003 11 12 & 17:20:32 & 300 & 2452956.224 & 14.481 & 0.013 \\
2003 11 12 & 17:39:39 & 300 & 2452956.238 & 14.494 & 0.013 \\
2003 11 12 & 17:58:25 & 300 & 2452956.251 & 14.475 & 0.012 \\
2003 11 12 & 18:17:22 & 300 & 2452956.264 & 14.474 & 0.014 \\
2003 11 12 & 18:36:17 & 300 & 2452956.277 & 14.514 & 0.015 \\
2003 11 12 & 18:55:01 & 300 & 2452956.290 & 14.474 & 0.014 \\
2003 11 12 & 19:13:47 & 300 & 2452956.303 & 14.476 & 0.015 \\
2003 11 12 & 19:32:46 & 300 & 2452956.316 & 14.482 & 0.016 \\
2003 11 12 & 19:51:35 & 300 & 2452956.329 & 14.494 & 0.016 \\
2003 11 12 & 20:10:47 & 300 & 2452956.343 & 14.526 & 0.017 \\
2003 11 12 & 20:29:56 & 300 & 2452956.356 & 14.570 & 0.017 \\
2003 11 12 & 20:48:50 & 300 & 2452956.369 & 14.517 & 0.024 \\
2003 11 12 & 21:07:33 & 300 & 2452956.382 & 14.574 & 0.017 \\
2003 11 12 & 21:26:17 & 300 & 2452956.395 & 14.602 & 0.019 \\
2003 11 12 & 21:45:15 & 300 & 2452956.408 & 14.603 & 0.021 \\
2003 11 14 & 16:01:49 & 300 & 2452958.170 & 14.532 & 0.014 \\
2003 11 14 & 16:20:33 & 300 & 2452958.183 & 14.555 & 0.017 \\
2003 11 14 & 16:39:19 & 300 & 2452958.196 & 14.559 & 0.014 \\
2003 11 14 & 16:58:17 & 300 & 2452958.209 & 14.546 & 0.014 \\
2003 11 14 & 17:17:03 & 300 & 2452958.222 & 14.534 & 0.012 \\
2003 11 14 & 17:36:01 & 300 & 2452958.235 & 14.525 & 0.014 \\
2003 11 14 & 17:54:44 & 300 & 2452958.248 & 14.551 & 0.011 \\
2003 11 14 & 18:13:44 & 300 & 2452958.261 & 14.530 & 0.012 \\
2003 11 14 & 18:33:09 & 300 & 2452958.275 & 14.550 & 0.014 \\
2003 11 14 & 18:52:11 & 300 & 2452958.288 & 14.520 & 0.011 \\
2003 11 14 & 19:10:58 & 300 & 2452958.301 & 14.508 & 0.010 \\
2003 11 14 & 19:29:53 & 300 & 2452958.314 & 14.539 & 0.010 \\
2003 11 14 & 19:48:41 & 300 & 2452958.327 & 14.495 & 0.010 \\
2003 11 14 & 20:08:04 & 300 & 2452958.341 & 14.480 & 0.011 \\
2003 11 14 & 20:26:49 & 300 & 2452958.354 & 14.468 & 0.010 \\
2003 11 14 & 20:45:50 & 300 & 2452958.367 & 14.442 & 0.010 \\
2003 11 14 & 21:04:46 & 300 & 2452958.380 & 14.452 & 0.010 \\
2003 11 14 & 21:23:51 & 300 & 2452958.393 & 14.460 & 0.009 \\
2003 11 14 & 21:43:00 & 300 & 2452958.407 & 14.447 & 0.009 \\
2003 11 14 & 22:01:56 & 300 & 2452958.420 & 14.432 & 0.024 \\
2003 11 15 & 15:25:35 & 300 & 2452959.145 & 14.317 & 0.010 \\
2003 11 15 & 15:47:07 & 300 & 2452959.160 & 14.332 & 0.009 \\
2003 11 15 & 15:52:51 & 300 & 2452959.163 & 14.318 & 0.009 \\
2003 11 15 & 16:15:41 & 300 & 2452959.179 & 14.347 & 0.009 \\
2003 11 15 & 16:34:23 & 300 & 2452959.192 & 14.351 & 0.009 \\
2003 11 15 & 16:53:09 & 300 & 2452959.205 & 14.384 & 0.009 \\
2003 11 15 & 17:12:05 & 300 & 2452959.219 & 14.366 & 0.009 \\
2003 11 15 & 17:31:00 & 300 & 2452959.232 & 14.378 & 0.009 \\
2003 11 15 & 17:50:05 & 300 & 2452959.245 & 14.384 & 0.010 \\
2003 11 15 & 18:08:50 & 300 & 2452959.258 & 14.357 & 0.009 \\
2003 11 15 & 18:27:45 & 300 & 2452959.271 & 14.368 & 0.010 \\
2003 11 15 & 18:46:49 & 300 & 2452959.284 & 14.357 & 0.009 \\
2003 11 15 & 19:05:34 & 300 & 2452959.297 & 14.363 & 0.009 \\
2003 11 15 & 19:24:27 & 300 & 2452959.310 & 14.347 & 0.010 \\
2003 11 15 & 19:43:10 & 300 & 2452959.323 & 14.348 & 0.009 \\
2003 11 15 & 20:02:26 & 300 & 2452959.337 & 14.347 & 0.009 \\
2003 11 15 & 20:21:31 & 300 & 2452959.350 & 14.341 & 0.008 \\
2003 11 15 & 20:40:28 & 300 & 2452959.363 & 14.313 & 0.009 \\
2003 11 15 & 20:59:30 & 300 & 2452959.376 & 14.328 & 0.009 \\
2003 11 15 & 21:18:26 & 300 & 2452959.389 & 14.334 & 0.009 \\
2003 11 15 & 21:37:23 & 300 & 2452959.403 & 14.323 & 0.009 \\
2003 11 15 & 21:56:14 & 300 & 2452959.416 & 14.338 & 0.013 \\
2003 11 16 & 15:05:24 & 300 & 2452960.131 & 14.236 & 0.007 \\
2003 11 16 & 15:26:57 & 300 & 2452960.146 & 14.253 & 0.007 \\
2003 11 16 & 15:46:08 & 300 & 2452960.159 & 14.254 & 0.007 \\
2003 11 16 & 16:05:08 & 300 & 2452960.172 & 14.254 & 0.007 \\
2003 11 16 & 16:24:16 & 300 & 2452960.185 & 14.295 & 0.008 \\
2003 11 16 & 16:43:05 & 300 & 2452960.198 & 14.313 & 0.008 \\
2003 11 16 & 17:01:50 & 300 & 2452960.211 & 14.329 & 0.009 \\
2003 11 16 & 17:20:50 & 300 & 2452960.225 & 14.318 & 0.008 \\
2003 11 16 & 17:39:47 & 300 & 2452960.238 & 14.315 & 0.013 \\
2003 11 16 & 17:52:19 & 300 & 2452960.246 & 14.305 & 0.013 \\
2003 11 16 & 18:11:03 & 300 & 2452960.259 & 14.297 & 0.015 \\
2003 11 16 & 18:30:03 & 300 & 2452960.273 & 14.283 & 0.022 \\
2003 11 16 & 18:48:49 & 300 & 2452960.286 & 14.302 & 0.017 \\
2003 11 16 & 19:07:47 & 300 & 2452960.299 & 14.280 & 0.008 \\
2003 11 16 & 19:20:16 & 300 & 2452960.308 & 14.283 & 0.008 \\
2003 11 16 & 19:39:15 & 300 & 2452960.321 & 14.303 & 0.008 \\
2003 11 16 & 19:58:14 & 300 & 2452960.334 & 14.316 & 0.008 \\
2003 11 16 & 20:17:19 & 300 & 2452960.347 & 14.295 & 0.008 \\
2003 11 16 & 20:36:44 & 300 & 2452960.361 & 14.288 & 0.009 \\
2003 11 16 & 20:49:16 & 300 & 2452960.369 & 14.294 & 0.010 \\
2003 11 16 & 21:08:17 & 300 & 2452960.382 & 14.338 & 0.017 \\
2003 11 18 & 15:26:36 & 300 & 2452962.145 & 14.351 & 0.016 \\
2003 11 18 & 15:48:54 & 300 & 2452962.161 & 14.350 & 0.016 \\
2003 11 18 & 16:07:39 & 300 & 2452962.174 & 14.371 & 0.030 \\
\enddata
\end{deluxetable}
\clearpage
\begin{deluxetable}{cccccc}
\tablecaption{Observational log and results in the BATC $k$ band in Period 1.\label{T4}}
\tablewidth{0pt}
\tablehead{
\colhead{Obs. Date} & \colhead{Obs. Time} & \colhead{Exp. Time} & \colhead{JD} & \colhead{$k$} & \colhead{$k_{\rm err}$} \\
\colhead{(yyyy mm dd)} & \colhead{(hh:mm:ss)} &  \colhead{(s)} & & \colhead{(mag)} & \colhead{(mag)}
}
\startdata
2003 11 08 & 16:33:17 & 100 & 2452952.190 & 13.869 & 0.021 \\
2003 11 08 & 18:11:18 & 300 & 2452952.260 & 13.882 & 0.010 \\
2003 11 08 & 18:46:28 & 300 & 2452952.284 & 13.879 & 0.010 \\
2003 11 08 & 19:21:27 & 300 & 2452952.308 & 13.867 & 0.010 \\
2003 11 08 & 19:48:44 & 300 & 2452952.327 & 13.869 & 0.010 \\
2003 11 08 & 20:20:24 & 300 & 2452952.349 & 13.862 & 0.015 \\
2003 11 08 & 20:47:55 & 300 & 2452952.368 & 13.813 & 0.025 \\
2003 11 08 & 21:15:07 & 300 & 2452952.387 & 13.849 & 0.025 \\
2003 11 08 & 21:42:28 & 300 & 2452952.406 & 13.884 & 0.042 \\
2003 11 12 & 16:49:24 & 300 & 2452956.203 & 14.035 & 0.012 \\
2003 11 12 & 17:08:07 & 300 & 2452956.216 & 14.031 & 0.012 \\
2003 11 12 & 17:26:45 & 300 & 2452956.229 & 14.051 & 0.012 \\
2003 11 12 & 17:45:55 & 300 & 2452956.242 & 14.029 & 0.011 \\
2003 11 12 & 18:04:41 & 300 & 2452956.255 & 14.052 & 0.011 \\
2003 11 12 & 18:23:35 & 300 & 2452956.268 & 14.057 & 0.012 \\
2003 11 12 & 18:42:29 & 300 & 2452956.281 & 14.068 & 0.013 \\
2003 11 12 & 19:01:19 & 300 & 2452956.294 & 14.013 & 0.012 \\
2003 11 12 & 19:20:14 & 300 & 2452956.308 & 14.024 & 0.013 \\
2003 11 12 & 19:39:04 & 300 & 2452956.320 & 14.045 & 0.013 \\
2003 11 12 & 19:57:58 & 300 & 2452956.334 & 14.083 & 0.014 \\
2003 11 12 & 20:17:15 & 300 & 2452956.347 & 14.087 & 0.015 \\
2003 11 12 & 20:36:10 & 300 & 2452956.360 & 14.128 & 0.015 \\
2003 11 12 & 20:55:05 & 300 & 2452956.373 & 14.136 & 0.017 \\
2003 11 12 & 21:13:46 & 300 & 2452956.386 & 14.184 & 0.016 \\
2003 11 12 & 21:32:44 & 300 & 2452956.400 & 14.207 & 0.017 \\
2003 11 12 & 21:51:40 & 300 & 2452956.413 & 14.199 & 0.020 \\
2003 11 14 & 16:08:07 & 300 & 2452958.174 & 14.161 & 0.014 \\
2003 11 14 & 16:26:47 & 300 & 2452958.187 & 14.152 & 0.014 \\
2003 11 14 & 16:45:46 & 300 & 2452958.200 & 14.149 & 0.014 \\
2003 11 14 & 17:04:32 & 300 & 2452958.213 & 14.150 & 0.013 \\
2003 11 14 & 17:23:34 & 300 & 2452958.226 & 14.152 & 0.011 \\
2003 11 14 & 17:42:14 & 300 & 2452958.239 & 14.207 & 0.012 \\
2003 11 14 & 18:01:14 & 300 & 2452958.253 & 14.166 & 0.011 \\
2003 11 14 & 18:20:12 & 300 & 2452958.266 & 14.162 & 0.012 \\
2003 11 14 & 18:39:25 & 300 & 2452958.279 & 14.159 & 0.012 \\
2003 11 14 & 18:58:25 & 300 & 2452958.292 & 14.162 & 0.011 \\
2003 11 14 & 19:17:15 & 300 & 2452958.305 & 14.123 & 0.011 \\
2003 11 14 & 19:36:05 & 300 & 2452958.319 & 14.118 & 0.009 \\
2003 11 14 & 19:55:12 & 300 & 2452958.332 & 14.133 & 0.010 \\
2003 11 14 & 20:14:18 & 300 & 2452958.345 & 14.114 & 0.011 \\
2003 11 14 & 20:33:08 & 300 & 2452958.358 & 14.093 & 0.010 \\
2003 11 14 & 20:52:04 & 300 & 2452958.371 & 14.088 & 0.010 \\
2003 11 14 & 21:11:11 & 300 & 2452958.384 & 14.117 & 0.010 \\
2003 11 14 & 21:30:06 & 300 & 2452958.398 & 14.108 & 0.010 \\
2003 11 14 & 21:49:14 & 300 & 2452958.411 & 14.079 & 0.010 \\
2003 11 15 & 15:33:15 & 300 & 2452959.150 & 13.928 & 0.009 \\
2003 11 15 & 16:03:03 & 300 & 2452959.170 & 13.941 & 0.009 \\
2003 11 15 & 16:21:53 & 300 & 2452959.184 & 13.955 & 0.009 \\
2003 11 15 & 16:40:40 & 300 & 2452959.197 & 13.973 & 0.009 \\
2003 11 15 & 16:59:23 & 300 & 2452959.210 & 13.997 & 0.009 \\
2003 11 15 & 17:18:20 & 300 & 2452959.223 & 13.974 & 0.010 \\
2003 11 15 & 17:37:22 & 300 & 2452959.236 & 13.995 & 0.010 \\
2003 11 15 & 17:56:21 & 300 & 2452959.249 & 13.999 & 0.010 \\
2003 11 15 & 18:15:03 & 300 & 2452959.262 & 13.995 & 0.009 \\
2003 11 15 & 18:34:12 & 300 & 2452959.275 & 14.000 & 0.009 \\
2003 11 15 & 18:53:04 & 300 & 2452959.289 & 13.982 & 0.009 \\
2003 11 15 & 19:12:02 & 300 & 2452959.302 & 13.988 & 0.009 \\
2003 11 15 & 19:30:42 & 300 & 2452959.315 & 13.990 & 0.010 \\
2003 11 15 & 19:49:37 & 300 & 2452959.328 & 13.999 & 0.010 \\
2003 11 15 & 20:08:58 & 300 & 2452959.341 & 13.987 & 0.009 \\
2003 11 15 & 20:27:59 & 300 & 2452959.354 & 13.968 & 0.009 \\
2003 11 15 & 20:46:42 & 300 & 2452959.368 & 13.962 & 0.009 \\
2003 11 15 & 21:05:59 & 300 & 2452959.381 & 13.973 & 0.009 \\
2003 11 15 & 21:24:51 & 300 & 2452959.394 & 13.987 & 0.009 \\
2003 11 15 & 21:43:46 & 300 & 2452959.407 & 13.956 & 0.009 \\
2003 11 15 & 22:02:28 & 300 & 2452959.420 & 13.953 & 0.022 \\
2003 11 16 & 15:33:25 & 300 & 2452960.150 & 13.889 & 0.008 \\
2003 11 16 & 15:52:23 & 300 & 2452960.163 & 13.894 & 0.007 \\
2003 11 16 & 16:11:34 & 300 & 2452960.176 & 13.931 & 0.009 \\
2003 11 16 & 16:30:32 & 300 & 2452960.190 & 13.934 & 0.008 \\
2003 11 16 & 16:49:18 & 300 & 2452960.203 & 13.932 & 0.008 \\
2003 11 16 & 17:08:05 & 300 & 2452960.216 & 13.927 & 0.009 \\
2003 11 16 & 17:27:04 & 300 & 2452960.229 & 13.943 & 0.010 \\
2003 11 16 & 17:58:33 & 300 & 2452960.251 & 13.920 & 0.012 \\
2003 11 16 & 18:17:33 & 300 & 2452960.264 & 13.912 & 0.024 \\
2003 11 16 & 18:36:18 & 300 & 2452960.277 & 13.864 & 0.027 \\
2003 11 16 & 18:55:07 & 300 & 2452960.290 & 13.882 & 0.015 \\
2003 11 16 & 19:26:31 & 300 & 2452960.312 & 13.908 & 0.008 \\
2003 11 16 & 19:45:30 & 300 & 2452960.325 & 13.927 & 0.008 \\
2003 11 16 & 20:04:36 & 300 & 2452960.338 & 13.923 & 0.008 \\
2003 11 16 & 20:23:46 & 300 & 2452960.352 & 13.904 & 0.009 \\
2003 11 16 & 20:55:33 & 300 & 2452960.374 & 13.958 & 0.013 \\
2003 11 18 & 15:13:55 & 300 & 2452962.136 & 13.969 & 0.014 \\
2003 11 18 & 15:32:58 & 300 & 2452962.150 & 13.967 & 0.014 \\
\enddata
\end{deluxetable}
\clearpage
\begin{deluxetable}{cccccc}
\tablecaption{Observational log and results in the BATC $e$ band in Period 2.\label{T5}}
\tablewidth{0pt}
\tablehead{
\colhead{Obs. Date} & \colhead{Obs. Time} & \colhead{Exp. Time} & \colhead{JD} & \colhead{$e$} & \colhead{$e_{\rm err}$} \\
\colhead{(yyyy mm dd)} & \colhead{(hh:mm:ss)} &  \colhead{(s)} & & \colhead{(mag)} & \colhead{(mag)}
}
\startdata
2003 12 30 & 12:31:07 & 300 & 2453004.023 & 14.535 & 0.015 \\
2003 12 30 & 12:42:56 & 300 & 2453004.032 & 14.505 & 0.014 \\
2003 12 30 & 12:59:40 & 300 & 2453004.043 & 14.506 & 0.013 \\
2003 12 30 & 13:16:10 & 300 & 2453004.055 & 14.514 & 0.014 \\
2003 12 30 & 13:32:42 & 300 & 2453004.066 & 14.534 & 0.014 \\
2003 12 30 & 13:53:08 & 300 & 2453004.080 & 14.519 & 0.014 \\
2003 12 30 & 14:12:09 & 300 & 2453004.094 & 14.545 & 0.013 \\
2003 12 30 & 14:28:39 & 300 & 2453004.105 & 14.505 & 0.013 \\
2003 12 30 & 14:49:45 & 300 & 2453004.120 & 14.524 & 0.012 \\
2003 12 30 & 15:06:09 & 300 & 2453004.131 & 14.515 & 0.013 \\
2003 12 30 & 15:22:46 & 300 & 2453004.142 & 14.492 & 0.013 \\
2003 12 30 & 15:39:17 & 300 & 2453004.154 & 14.516 & 0.016 \\
2003 12 30 & 15:56:00 & 300 & 2453004.166 & 14.444 & 0.012 \\
2003 12 30 & 16:12:17 & 300 & 2453004.177 & 14.427 & 0.012 \\
2003 12 30 & 16:28:42 & 300 & 2453004.188 & 14.436 & 0.012 \\
2003 12 30 & 16:45:08 & 300 & 2453004.200 & 14.417 & 0.012 \\
2003 12 30 & 17:01:36 & 300 & 2453004.211 & 14.414 & 0.011 \\
2003 12 30 & 17:18:16 & 300 & 2453004.223 & 14.405 & 0.011 \\
2003 12 30 & 17:34:50 & 300 & 2453004.234 & 14.401 & 0.010 \\
2003 12 30 & 17:51:22 & 300 & 2453004.246 & 14.434 & 0.010 \\
2003 12 30 & 18:11:39 & 300 & 2453004.260 & 14.396 & 0.009 \\
2003 12 30 & 18:28:22 & 300 & 2453004.271 & 14.406 & 0.009 \\
2003 12 30 & 18:44:52 & 300 & 2453004.283 & 14.400 & 0.010 \\
2003 12 30 & 19:01:11 & 300 & 2453004.294 & 14.394 & 0.009 \\
2003 12 30 & 19:17:39 & 300 & 2453004.306 & 14.437 & 0.010 \\
2003 12 30 & 19:34:08 & 300 & 2453004.317 & 14.415 & 0.010 \\
2003 12 30 & 19:50:37 & 300 & 2453004.329 & 14.416 & 0.010 \\
2003 12 30 & 20:07:34 & 300 & 2453004.340 & 14.410 & 0.010 \\
2003 12 30 & 20:26:57 & 300 & 2453004.354 & 14.400 & 0.009 \\
2003 12 30 & 20:43:33 & 300 & 2453004.365 & 14.374 & 0.009 \\
2003 12 30 & 20:59:53 & 300 & 2453004.377 & 14.363 & 0.010 \\
2003 12 30 & 21:16:36 & 300 & 2453004.388 & 14.373 & 0.010 \\
2003 12 30 & 21:33:15 & 300 & 2453004.400 & 14.415 & 0.012 \\
2003 12 30 & 21:49:50 & 300 & 2453004.411 & 14.418 & 0.010 \\
2003 12 30 & 22:06:23 & 300 & 2453004.423 & 14.412 & 0.011 \\
2003 12 31 & 13:16:36 & 300 & 2453005.055 & 14.288 & 0.011 \\
2003 12 31 & 13:33:01 & 300 & 2453005.066 & 14.315 & 0.011 \\
2003 12 31 & 13:49:36 & 300 & 2453005.078 & 14.305 & 0.011 \\
2003 12 31 & 14:05:59 & 300 & 2453005.089 & 14.321 & 0.011 \\
2003 12 31 & 14:22:26 & 300 & 2453005.101 & 14.323 & 0.012 \\
2003 12 31 & 14:38:55 & 300 & 2453005.112 & 14.331 & 0.012 \\
2003 12 31 & 14:56:20 & 300 & 2453005.124 & 14.369 & 0.012 \\
2003 12 31 & 15:12:50 & 300 & 2453005.136 & 14.375 & 0.014 \\
2003 12 31 & 15:29:04 & 300 & 2453005.147 & 14.382 & 0.012 \\
2003 12 31 & 15:45:18 & 300 & 2453005.158 & 14.372 & 0.012 \\
2003 12 31 & 16:01:42 & 300 & 2453005.170 & 14.348 & 0.012 \\
2003 12 31 & 16:18:27 & 300 & 2453005.181 & 14.354 & 0.012 \\
2003 12 31 & 16:36:37 & 300 & 2453005.194 & 14.310 & 0.011 \\
2003 12 31 & 16:53:03 & 300 & 2453005.205 & 14.333 & 0.011 \\
2003 12 31 & 17:09:18 & 300 & 2453005.216 & 14.352 & 0.011 \\
2003 12 31 & 17:42:14 & 300 & 2453005.239 & 14.316 & 0.012 \\
2003 12 31 & 17:59:11 & 300 & 2453005.251 & 14.313 & 0.012 \\
2003 12 31 & 18:42:17 & 300 & 2453005.281 & 14.278 & 0.011 \\
2003 12 31 & 18:59:28 & 300 & 2453005.293 & 14.281 & 0.012 \\
2003 12 31 & 19:19:37 & 300 & 2453005.307 & 14.274 & 0.011 \\
2003 12 31 & 19:36:02 & 300 & 2453005.318 & 14.261 & 0.010 \\
2003 12 31 & 19:54:17 & 300 & 2453005.331 & 14.231 & 0.009 \\
2003 12 31 & 20:10:59 & 300 & 2453005.343 & 14.201 & 0.009 \\
2003 12 31 & 20:27:25 & 300 & 2453005.354 & 14.180 & 0.009 \\
2003 12 31 & 20:43:51 & 300 & 2453005.366 & 14.187 & 0.009 \\
2003 12 31 & 21:00:31 & 300 & 2453005.377 & 14.111 & 0.009 \\
2003 12 31 & 21:22:06 & 300 & 2453005.392 & 14.108 & 0.008 \\
2003 12 31 & 21:38:20 & 300 & 2453005.403 & 14.071 & 0.008 \\
2003 12 31 & 21:57:07 & 300 & 2453005.416 & 14.036 & 0.008 \\
2004 01 01 & 12:36:21 & 300 & 2453006.027 & 14.325 & 0.009 \\
2004 01 01 & 12:52:49 & 300 & 2453006.038 & 14.303 & 0.009 \\
2004 01 01 & 13:09:33 & 300 & 2453006.050 & 14.294 & 0.009 \\
2004 01 01 & 13:26:34 & 300 & 2453006.062 & 14.266 & 0.009 \\
2004 01 01 & 13:54:33 & 300 & 2453006.081 & 14.272 & 0.009 \\
2004 01 01 & 14:11:39 & 300 & 2453006.093 & 14.284 & 0.009 \\
2004 01 01 & 14:29:25 & 300 & 2453006.105 & 14.268 & 0.008 \\
2004 01 01 & 14:46:18 & 300 & 2453006.117 & 14.277 & 0.008 \\
2004 01 01 & 15:13:28 & 300 & 2453006.136 & 14.290 & 0.008 \\
2004 01 01 & 15:29:49 & 300 & 2453006.147 & 14.312 & 0.009 \\
2004 01 01 & 15:46:07 & 300 & 2453006.159 & 14.340 & 0.009 \\
2004 01 01 & 16:41:51 & 300 & 2453006.198 & 14.360 & 0.009 \\
2004 01 01 & 17:00:23 & 300 & 2453006.210 & 14.347 & 0.009 \\
2004 01 01 & 17:37:42 & 300 & 2453006.236 & 14.324 & 0.009 \\
2004 01 01 & 17:54:43 & 300 & 2453006.248 & 14.288 & 0.008 \\
2004 01 01 & 18:11:14 & 300 & 2453006.259 & 14.271 & 0.008 \\
2004 01 01 & 18:30:07 & 300 & 2453006.273 & 14.289 & 0.008 \\
2004 01 02 & 12:25:02 & 300 & 2453007.019 & 13.987 & 0.008 \\
2004 01 02 & 12:42:41 & 300 & 2453007.031 & 14.017 & 0.008 \\
2004 01 02 & 12:59:16 & 300 & 2453007.043 & 14.014 & 0.009 \\
2004 01 02 & 13:15:31 & 300 & 2453007.054 & 13.983 & 0.007 \\
2004 01 02 & 13:31:59 & 300 & 2453007.066 & 13.992 & 0.007 \\
2004 01 02 & 13:48:17 & 300 & 2453007.077 & 13.988 & 0.007 \\
2004 01 02 & 14:04:59 & 300 & 2453007.089 & 13.992 & 0.007 \\
2004 01 02 & 14:21:37 & 300 & 2453007.100 & 14.006 & 0.008 \\
2004 01 02 & 14:37:55 & 300 & 2453007.111 & 14.041 & 0.008 \\
2004 01 02 & 14:54:20 & 300 & 2453007.123 & 14.031 & 0.007 \\
2004 01 02 & 15:29:14 & 300 & 2453007.147 & 14.028 & 0.007 \\
2004 01 02 & 15:45:25 & 300 & 2453007.158 & 14.031 & 0.007 \\
2004 01 02 & 16:01:40 & 300 & 2453007.170 & 14.027 & 0.007 \\
2004 01 02 & 16:19:13 & 300 & 2453007.182 & 14.015 & 0.007 \\
2004 01 02 & 16:35:30 & 300 & 2453007.193 & 14.019 & 0.007 \\
2004 01 02 & 16:52:17 & 300 & 2453007.205 & 14.031 & 0.007 \\
2004 01 02 & 17:08:34 & 300 & 2453007.216 & 14.034 & 0.007 \\
2004 01 02 & 17:25:09 & 300 & 2453007.228 & 14.038 & 0.007 \\
2004 01 02 & 17:41:39 & 300 & 2453007.239 & 14.045 & 0.007 \\
2004 01 02 & 18:01:28 & 300 & 2453007.253 & 14.038 & 0.007 \\
2004 01 02 & 18:18:19 & 300 & 2453007.264 & 14.019 & 0.007 \\
2004 01 02 & 18:35:03 & 300 & 2453007.276 & 14.004 & 0.007 \\
2004 01 02 & 18:51:36 & 300 & 2453007.288 & 13.994 & 0.007 \\
2004 01 02 & 19:08:12 & 300 & 2453007.299 & 14.002 & 0.007 \\
2004 01 02 & 19:25:11 & 300 & 2453007.311 & 14.005 & 0.007 \\
2004 01 03 & 16:05:05 & 300 & 2453008.172 & 14.015 & 0.009 \\
2004 01 03 & 16:21:41 & 300 & 2453008.184 & 13.999 & 0.009 \\
2004 01 03 & 16:38:03 & 300 & 2453008.195 & 14.002 & 0.009 \\
2004 01 03 & 16:54:46 & 300 & 2453008.206 & 13.984 & 0.009 \\
2004 01 03 & 17:11:18 & 300 & 2453008.218 & 13.996 & 0.009 \\
2004 01 03 & 17:27:37 & 300 & 2453008.229 & 13.981 & 0.009 \\
2004 01 03 & 17:44:25 & 300 & 2453008.241 & 13.964 & 0.009 \\
2004 01 03 & 18:01:43 & 300 & 2453008.253 & 13.972 & 0.009 \\
2004 01 03 & 18:18:13 & 300 & 2453008.264 & 13.965 & 0.009 \\
2004 01 03 & 18:34:42 & 300 & 2453008.276 & 13.986 & 0.009 \\
2004 01 03 & 18:51:14 & 300 & 2453008.287 & 13.977 & 0.009 \\
2004 01 03 & 19:07:52 & 300 & 2453008.299 & 13.960 & 0.009 \\
2004 01 03 & 19:24:43 & 300 & 2453008.311 & 13.973 & 0.009 \\
2004 01 03 & 19:41:02 & 300 & 2453008.322 & 13.949 & 0.009 \\
2004 01 03 & 19:57:38 & 300 & 2453008.333 & 13.954 & 0.008 \\
2004 01 03 & 20:14:43 & 300 & 2453008.345 & 13.959 & 0.008 \\
2004 01 03 & 20:31:59 & 300 & 2453008.357 & 13.965 & 0.008 \\
2004 01 03 & 20:53:05 & 300 & 2453008.372 & 13.969 & 0.008 \\
2004 01 04 & 12:10:37 & 300 & 2453009.009 & 14.057 & 0.010 \\
2004 01 04 & 12:27:18 & 300 & 2453009.021 & 14.064 & 0.010 \\
2004 01 04 & 12:49:37 & 300 & 2453009.036 & 14.060 & 0.010 \\
2004 01 04 & 13:06:02 & 300 & 2453009.048 & 14.026 & 0.009 \\
2004 01 04 & 13:22:18 & 300 & 2453009.059 & 14.023 & 0.009 \\
2004 01 04 & 15:57:09 & 300 & 2453009.166 & 13.966 & 0.008 \\
2004 01 04 & 16:13:39 & 300 & 2453009.178 & 13.974 & 0.008 \\
2004 01 04 & 16:29:57 & 300 & 2453009.189 & 13.956 & 0.008 \\
2004 01 04 & 16:46:23 & 300 & 2453009.201 & 13.947 & 0.008 \\
2004 01 04 & 17:04:24 & 300 & 2453009.213 & 13.939 & 0.008 \\
2004 01 04 & 17:20:49 & 300 & 2453009.225 & 13.926 & 0.008 \\
2004 01 04 & 17:38:12 & 300 & 2453009.237 & 13.916 & 0.008 \\
2004 01 04 & 17:54:44 & 300 & 2453009.248 & 13.939 & 0.008 \\
2004 01 04 & 18:11:11 & 300 & 2453009.259 & 13.898 & 0.008 \\
2004 01 04 & 18:27:30 & 300 & 2453009.271 & 13.884 & 0.008 \\
2004 01 04 & 18:46:06 & 300 & 2453009.284 & 13.878 & 0.008 \\
2004 01 04 & 19:02:31 & 300 & 2453009.295 & 13.854 & 0.007 \\
2004 01 04 & 19:19:24 & 300 & 2453009.307 & 13.838 & 0.007 \\
2004 01 04 & 19:35:49 & 300 & 2453009.318 & 13.855 & 0.008 \\
2004 01 04 & 19:52:27 & 300 & 2453009.330 & 13.852 & 0.008 \\
2004 01 04 & 20:10:56 & 300 & 2453009.343 & 13.811 & 0.007 \\
2004 01 04 & 20:27:21 & 300 & 2453009.354 & 13.814 & 0.007 \\
2004 01 04 & 20:43:50 & 300 & 2453009.366 & 13.809 & 0.007 \\
2004 01 04 & 21:00:48 & 300 & 2453009.377 & 13.791 & 0.007 \\
2004 01 04 & 21:17:11 & 300 & 2453009.389 & 13.768 & 0.007 \\
2004 01 04 & 21:34:46 & 300 & 2453009.401 & 13.777 & 0.006 \\
2004 01 05 & 13:36:53 & 300 & 2453010.069 & 13.854 & 0.009 \\
2004 01 05 & 13:57:26 & 300 & 2453010.083 & 13.851 & 0.009 \\
2004 01 05 & 14:14:16 & 300 & 2453010.095 & 13.853 & 0.010 \\
2004 01 05 & 14:30:43 & 300 & 2453010.106 & 13.857 & 0.013 \\
\enddata
\end{deluxetable}
\clearpage
\begin{deluxetable}{cccccc}
\tablecaption{Observational log and results in the BATC $f$ band in Period 2.\label{T6}}
\tablewidth{0pt}
\tablehead{
\colhead{Obs. Date} & \colhead{Obs. Time} & \colhead{Exp. Time} & \colhead{JD} & \colhead{$f$} & \colhead{$f_{\rm err}$} \\
\colhead{(yyyy mm dd)} & \colhead{(hh:mm:ss)} &  \colhead{(s)} & & \colhead{(mag)} & \colhead{(mag)}
}
\startdata
2003 12 30 & 12:49:12 & 300 & 2453004.036 & 14.430 & 0.012 \\
2003 12 30 & 13:05:58 & 300 & 2453004.047 & 14.445 & 0.012 \\
2003 12 30 & 13:22:14 & 300 & 2453004.059 & 14.436 & 0.012 \\
2003 12 30 & 13:38:47 & 300 & 2453004.070 & 14.448 & 0.011 \\
2003 12 30 & 13:59:29 & 300 & 2453004.085 & 14.426 & 0.012 \\
2003 12 30 & 14:18:16 & 300 & 2453004.098 & 14.433 & 0.011 \\
2003 12 30 & 14:34:44 & 300 & 2453004.109 & 14.425 & 0.011 \\
2003 12 30 & 14:56:00 & 300 & 2453004.124 & 14.416 & 0.010 \\
2003 12 30 & 15:12:26 & 300 & 2453004.135 & 14.406 & 0.011 \\
2003 12 30 & 15:28:52 & 300 & 2453004.147 & 14.425 & 0.012 \\
2003 12 30 & 15:45:31 & 300 & 2453004.158 & 14.381 & 0.012 \\
2003 12 30 & 16:02:08 & 300 & 2453004.170 & 14.347 & 0.010 \\
2003 12 30 & 16:18:23 & 300 & 2453004.181 & 14.359 & 0.010 \\
2003 12 30 & 16:34:50 & 300 & 2453004.193 & 14.361 & 0.010 \\
2003 12 30 & 16:51:27 & 300 & 2453004.204 & 14.334 & 0.010 \\
2003 12 30 & 17:07:55 & 300 & 2453004.216 & 14.351 & 0.009 \\
2003 12 30 & 17:24:32 & 300 & 2453004.227 & 14.324 & 0.009 \\
2003 12 30 & 17:40:55 & 300 & 2453004.238 & 14.326 & 0.008 \\
2003 12 30 & 17:57:27 & 300 & 2453004.250 & 14.321 & 0.008 \\
2003 12 30 & 18:17:44 & 300 & 2453004.264 & 14.327 & 0.008 \\
2003 12 30 & 18:34:29 & 300 & 2453004.276 & 14.331 & 0.008 \\
2003 12 30 & 18:50:58 & 300 & 2453004.287 & 14.322 & 0.008 \\
2003 12 30 & 19:07:16 & 300 & 2453004.298 & 14.323 & 0.008 \\
2003 12 30 & 19:23:43 & 300 & 2453004.310 & 14.328 & 0.009 \\
2003 12 30 & 19:40:27 & 300 & 2453004.322 & 14.337 & 0.009 \\
2003 12 30 & 19:56:56 & 300 & 2453004.333 & 14.342 & 0.008 \\
2003 12 30 & 20:16:31 & 300 & 2453004.347 & 14.342 & 0.008 \\
2003 12 30 & 20:33:06 & 300 & 2453004.358 & 14.312 & 0.008 \\
2003 12 30 & 20:49:41 & 300 & 2453004.370 & 14.329 & 0.008 \\
2003 12 30 & 21:06:01 & 300 & 2453004.381 & 14.285 & 0.009 \\
2003 12 30 & 21:22:48 & 300 & 2453004.393 & 14.322 & 0.009 \\
2003 12 30 & 21:39:34 & 300 & 2453004.404 & 14.316 & 0.009 \\
2003 12 30 & 21:56:09 & 300 & 2453004.416 & 14.367 & 0.010 \\
2003 12 31 & 13:22:43 & 300 & 2453005.059 & 14.223 & 0.009 \\
2003 12 31 & 13:39:20 & 300 & 2453005.071 & 14.217 & 0.010 \\
2003 12 31 & 13:55:44 & 300 & 2453005.082 & 14.200 & 0.009 \\
2003 12 31 & 14:12:16 & 300 & 2453005.094 & 14.216 & 0.010 \\
2003 12 31 & 14:28:30 & 300 & 2453005.105 & 14.261 & 0.010 \\
2003 12 31 & 14:45:01 & 300 & 2453005.116 & 14.251 & 0.010 \\
2003 12 31 & 15:02:28 & 300 & 2453005.128 & 14.274 & 0.010 \\
2003 12 31 & 15:18:54 & 300 & 2453005.140 & 14.294 & 0.011 \\
2003 12 31 & 15:35:08 & 300 & 2453005.151 & 14.294 & 0.010 \\
2003 12 31 & 15:51:32 & 300 & 2453005.162 & 14.289 & 0.010 \\
2003 12 31 & 16:07:58 & 300 & 2453005.174 & 14.292 & 0.010 \\
2003 12 31 & 16:25:23 & 300 & 2453005.186 & 14.230 & 0.009 \\
2003 12 31 & 16:42:52 & 300 & 2453005.198 & 14.233 & 0.010 \\
2003 12 31 & 16:59:07 & 300 & 2453005.209 & 14.235 & 0.009 \\
2003 12 31 & 17:15:36 & 300 & 2453005.221 & 14.234 & 0.009 \\
2003 12 31 & 17:25:44 & 300 & 2453005.228 & 14.238 & 0.011 \\
2003 12 31 & 17:31:52 & 300 & 2453005.232 & 14.195 & 0.010 \\
2003 12 31 & 17:48:30 & 300 & 2453005.244 & 14.224 & 0.010 \\
2003 12 31 & 18:05:15 & 300 & 2453005.255 & 14.203 & 0.010 \\
2003 12 31 & 18:48:24 & 300 & 2453005.285 & 14.175 & 0.010 \\
2003 12 31 & 19:05:39 & 300 & 2453005.297 & 14.187 & 0.010 \\
2003 12 31 & 21:28:10 & 300 & 2453005.396 & 14.003 & 0.007 \\
2003 12 31 & 21:44:24 & 300 & 2453005.408 & 13.970 & 0.007 \\
2004 01 01 & 12:42:27 & 300 & 2453006.031 & 14.227 & 0.008 \\
2004 01 01 & 12:58:54 & 300 & 2453006.043 & 14.213 & 0.008 \\
2004 01 01 & 13:15:40 & 300 & 2453006.054 & 14.196 & 0.007 \\
2004 01 01 & 13:32:55 & 300 & 2453006.066 & 14.204 & 0.008 \\
2004 01 01 & 14:01:09 & 300 & 2453006.086 & 14.190 & 0.008 \\
2004 01 01 & 14:17:49 & 300 & 2453006.097 & 14.186 & 0.007 \\
2004 01 01 & 14:35:52 & 300 & 2453006.110 & 14.207 & 0.007 \\
2004 01 01 & 14:52:48 & 300 & 2453006.122 & 14.207 & 0.007 \\
2004 01 01 & 15:19:36 & 300 & 2453006.140 & 14.224 & 0.007 \\
2004 01 01 & 15:35:55 & 300 & 2453006.152 & 14.229 & 0.007 \\
2004 01 01 & 15:52:26 & 300 & 2453006.163 & 14.237 & 0.008 \\
2004 01 01 & 16:47:57 & 300 & 2453006.202 & 14.255 & 0.008 \\
2004 01 01 & 17:06:29 & 300 & 2453006.215 & 14.242 & 0.008 \\
2004 01 01 & 17:44:02 & 300 & 2453006.241 & 14.219 & 0.007 \\
2004 01 01 & 18:00:49 & 300 & 2453006.252 & 14.204 & 0.007 \\
2004 01 01 & 18:17:21 & 300 & 2453006.264 & 14.200 & 0.007 \\
2004 01 01 & 18:36:26 & 300 & 2453006.277 & 14.199 & 0.007 \\
2004 01 02 & 12:31:10 & 300 & 2453007.023 & 13.935 & 0.007 \\
2004 01 02 & 12:48:55 & 300 & 2453007.036 & 13.936 & 0.007 \\
2004 01 02 & 13:05:22 & 300 & 2453007.047 & 13.940 & 0.007 \\
2004 01 02 & 13:21:39 & 300 & 2453007.058 & 13.924 & 0.006 \\
2004 01 02 & 13:38:04 & 300 & 2453007.070 & 13.919 & 0.007 \\
2004 01 02 & 13:54:20 & 300 & 2453007.081 & 13.923 & 0.006 \\
2004 01 02 & 14:11:15 & 300 & 2453007.093 & 13.925 & 0.006 \\
2004 01 02 & 14:27:45 & 300 & 2453007.104 & 13.940 & 0.007 \\
2004 01 02 & 14:44:13 & 300 & 2453007.116 & 13.951 & 0.007 \\
2004 01 02 & 15:00:27 & 300 & 2453007.127 & 13.949 & 0.006 \\
2004 01 02 & 15:18:49 & 300 & 2453007.140 & 13.955 & 0.006 \\
2004 01 02 & 15:35:18 & 300 & 2453007.151 & 13.945 & 0.006 \\
2004 01 02 & 15:51:33 & 300 & 2453007.162 & 13.931 & 0.006 \\
2004 01 02 & 16:08:00 & 300 & 2453007.174 & 13.939 & 0.006 \\
2004 01 02 & 16:25:17 & 300 & 2453007.186 & 13.950 & 0.006 \\
2004 01 02 & 16:41:46 & 300 & 2453007.197 & 13.929 & 0.006 \\
2004 01 02 & 16:58:24 & 300 & 2453007.209 & 13.945 & 0.006 \\
2004 01 02 & 17:14:50 & 300 & 2453007.220 & 13.957 & 0.006 \\
2004 01 02 & 17:31:14 & 300 & 2453007.232 & 13.962 & 0.006 \\
2004 01 02 & 17:47:45 & 300 & 2453007.243 & 13.972 & 0.006 \\
2004 01 02 & 18:07:42 & 300 & 2453007.257 & 13.942 & 0.006 \\
2004 01 02 & 18:24:35 & 300 & 2453007.269 & 13.946 & 0.006 \\
2004 01 02 & 18:41:11 & 300 & 2453007.280 & 13.933 & 0.006 \\
2004 01 02 & 18:57:42 & 300 & 2453007.292 & 13.927 & 0.006 \\
2004 01 02 & 19:14:30 & 300 & 2453007.303 & 13.915 & 0.006 \\
2004 01 02 & 19:31:17 & 300 & 2453007.315 & 13.922 & 0.006 \\
2004 01 03 & 16:11:29 & 300 & 2453008.176 & 13.946 & 0.008 \\
2004 01 03 & 16:27:49 & 300 & 2453008.188 & 13.932 & 0.008 \\
2004 01 03 & 16:44:21 & 300 & 2453008.199 & 13.932 & 0.008 \\
2004 01 03 & 17:00:52 & 300 & 2453008.211 & 13.924 & 0.008 \\
2004 01 03 & 17:17:24 & 300 & 2453008.222 & 13.906 & 0.008 \\
2004 01 03 & 17:34:01 & 300 & 2453008.234 & 13.913 & 0.008 \\
2004 01 03 & 17:50:30 & 300 & 2453008.245 & 13.897 & 0.008 \\
2004 01 03 & 18:08:03 & 300 & 2453008.257 & 13.896 & 0.008 \\
2004 01 03 & 18:24:19 & 300 & 2453008.269 & 13.887 & 0.008 \\
2004 01 03 & 18:41:00 & 300 & 2453008.280 & 13.900 & 0.008 \\
2004 01 03 & 18:57:22 & 300 & 2453008.292 & 13.886 & 0.008 \\
2004 01 03 & 19:14:11 & 300 & 2453008.303 & 13.869 & 0.008 \\
2004 01 03 & 19:30:51 & 300 & 2453008.315 & 13.874 & 0.007 \\
2004 01 03 & 19:47:07 & 300 & 2453008.326 & 13.870 & 0.007 \\
2004 01 03 & 20:04:00 & 300 & 2453008.338 & 13.886 & 0.007 \\
2004 01 03 & 20:20:56 & 300 & 2453008.350 & 13.870 & 0.007 \\
2004 01 03 & 20:38:29 & 300 & 2453008.362 & 13.881 & 0.007 \\
2004 01 03 & 20:59:48 & 300 & 2453008.377 & 13.893 & 0.007 \\
2004 01 04 & 12:16:42 & 300 & 2453009.013 & 13.959 & 0.009 \\
2004 01 04 & 12:37:10 & 300 & 2453009.027 & 13.938 & 0.008 \\
2004 01 04 & 12:55:51 & 300 & 2453009.041 & 13.983 & 0.009 \\
2004 01 04 & 13:12:06 & 300 & 2453009.052 & 13.956 & 0.008 \\
2004 01 04 & 13:28:24 & 300 & 2453009.063 & 13.962 & 0.008 \\
2004 01 04 & 16:03:27 & 300 & 2453009.171 & 13.896 & 0.007 \\
2004 01 04 & 16:19:44 & 300 & 2453009.182 & 13.875 & 0.007 \\
2004 01 04 & 16:36:15 & 300 & 2453009.194 & 13.873 & 0.007 \\
2004 01 04 & 16:52:29 & 300 & 2453009.205 & 13.847 & 0.007 \\
2004 01 04 & 17:10:28 & 300 & 2453009.217 & 13.853 & 0.007 \\
2004 01 04 & 17:26:55 & 300 & 2453009.229 & 13.849 & 0.007 \\
2004 01 04 & 17:44:20 & 300 & 2453009.241 & 13.831 & 0.007 \\
2004 01 04 & 18:01:00 & 300 & 2453009.252 & 13.833 & 0.007 \\
2004 01 04 & 18:17:17 & 300 & 2453009.264 & 13.827 & 0.007 \\
2004 01 04 & 18:33:34 & 300 & 2453009.275 & 13.816 & 0.007 \\
2004 01 04 & 18:52:13 & 300 & 2453009.288 & 13.798 & 0.007 \\
2004 01 04 & 19:08:50 & 300 & 2453009.299 & 13.780 & 0.007 \\
2004 01 04 & 19:25:27 & 300 & 2453009.311 & 13.776 & 0.007 \\
2004 01 04 & 19:41:55 & 300 & 2453009.322 & 13.788 & 0.007 \\
2004 01 04 & 19:58:30 & 300 & 2453009.334 & 13.763 & 0.006 \\
2004 01 04 & 20:16:54 & 300 & 2453009.347 & 13.741 & 0.006 \\
2004 01 04 & 20:33:25 & 300 & 2453009.358 & 13.727 & 0.006 \\
2004 01 04 & 20:50:09 & 300 & 2453009.370 & 13.717 & 0.006 \\
2004 01 04 & 21:06:58 & 300 & 2453009.382 & 13.717 & 0.006 \\
2004 01 04 & 21:23:19 & 300 & 2453009.393 & 13.706 & 0.006 \\
2004 01 04 & 21:40:50 & 300 & 2453009.405 & 13.687 & 0.005 \\
2004 01 05 & 13:44:57 & 300 & 2453010.075 & 13.767 & 0.008 \\
2004 01 05 & 14:03:35 & 300 & 2453010.088 & 13.770 & 0.008 \\
2004 01 05 & 14:20:23 & 300 & 2453010.099 & 13.759 & 0.010 \\
2004 01 05 & 14:37:06 & 300 & 2453010.111 & 13.769 & 0.014 \\
\enddata
\end{deluxetable}
\clearpage
\begin{deluxetable}{cccccc}
\tablecaption{Observational log and results in the BATC $i$ band in Period 2.\label{T7}}
\tablewidth{0pt}
\tablehead{
\colhead{Obs. Date} & \colhead{Obs. Time} & \colhead{Exp. Time} & \colhead{JD} & \colhead{$i$} & \colhead{$i_{\rm err}$} \\
\colhead{(yyyy mm dd)} & \colhead{(hh:mm:ss)} &  \colhead{(s)} & & \colhead{(mag)} & \colhead{(mag)}
}
\startdata
2003 12 30 & 12:26:59 & 180 & 2453004.020 & 14.152 & 0.007 \\
2003 12 30 & 12:38:17 & 180 & 2453004.028 & 14.163 & 0.007 \\
2003 12 30 & 12:55:21 & 180 & 2453004.040 & 14.161 & 0.007 \\
2003 12 30 & 13:12:06 & 180 & 2453004.051 & 14.164 & 0.007 \\
2003 12 30 & 13:28:21 & 180 & 2453004.062 & 14.149 & 0.007 \\
2003 12 30 & 13:45:10 & 180 & 2453004.074 & 14.150 & 0.007 \\
2003 12 30 & 14:08:04 & 180 & 2453004.090 & 14.137 & 0.007 \\
2003 12 30 & 14:24:34 & 180 & 2453004.101 & 14.161 & 0.007 \\
2003 12 30 & 14:45:40 & 180 & 2453004.116 & 14.138 & 0.007 \\
2003 12 30 & 15:02:05 & 180 & 2453004.127 & 14.148 & 0.007 \\
2003 12 30 & 15:18:39 & 180 & 2453004.139 & 14.137 & 0.007 \\
2003 12 30 & 15:34:56 & 180 & 2453004.150 & 14.130 & 0.007 \\
2003 12 30 & 15:51:39 & 180 & 2453004.162 & 14.095 & 0.007 \\
2003 12 30 & 16:08:12 & 180 & 2453004.173 & 14.068 & 0.007 \\
2003 12 30 & 16:24:28 & 180 & 2453004.185 & 14.082 & 0.007 \\
2003 12 30 & 16:40:56 & 180 & 2453004.196 & 14.058 & 0.007 \\
2003 12 30 & 16:57:31 & 180 & 2453004.208 & 14.047 & 0.007 \\
2003 12 30 & 17:14:12 & 180 & 2453004.219 & 14.062 & 0.006 \\
2003 12 30 & 17:30:46 & 180 & 2453004.231 & 14.063 & 0.006 \\
2003 12 30 & 17:47:04 & 180 & 2453004.242 & 14.060 & 0.006 \\
2003 12 30 & 18:07:31 & 180 & 2453004.256 & 14.040 & 0.005 \\
2003 12 30 & 18:24:04 & 180 & 2453004.268 & 14.060 & 0.005 \\
2003 12 30 & 18:40:46 & 180 & 2453004.279 & 14.065 & 0.006 \\
2003 12 30 & 18:57:07 & 180 & 2453004.291 & 14.047 & 0.006 \\
2003 12 30 & 19:13:34 & 180 & 2453004.302 & 14.049 & 0.006 \\
2003 12 30 & 19:30:03 & 180 & 2453004.314 & 14.058 & 0.006 \\
2003 12 30 & 19:46:32 & 180 & 2453004.325 & 14.063 & 0.006 \\
2003 12 30 & 20:03:16 & 180 & 2453004.337 & 14.060 & 0.005 \\
2003 12 30 & 20:22:50 & 180 & 2453004.350 & 14.056 & 0.005 \\
2003 12 30 & 20:39:26 & 180 & 2453004.362 & 14.030 & 0.005 \\
2003 12 30 & 20:55:46 & 180 & 2453004.373 & 14.037 & 0.006 \\
2003 12 30 & 21:12:17 & 180 & 2453004.385 & 14.038 & 0.006 \\
2003 12 30 & 21:29:10 & 180 & 2453004.396 & 14.051 & 0.006 \\
2003 12 30 & 21:45:42 & 180 & 2453004.408 & 14.044 & 0.006 \\
2003 12 30 & 22:02:15 & 180 & 2453004.419 & 14.041 & 0.006 \\
2003 12 31 & 12:08:13 & 180 & 2453005.007 & 13.931 & 0.006 \\
2003 12 31 & 13:04:33 & 180 & 2453005.046 & 13.954 & 0.006 \\
2003 12 31 & 13:11:32 & 180 & 2453005.051 & 13.925 & 0.006 \\
2003 12 31 & 13:29:01 & 180 & 2453005.063 & 13.943 & 0.006 \\
2003 12 31 & 13:45:27 & 180 & 2453005.074 & 13.958 & 0.006 \\
2003 12 31 & 14:01:52 & 180 & 2453005.086 & 13.966 & 0.006 \\
2003 12 31 & 14:18:20 & 180 & 2453005.097 & 13.944 & 0.006 \\
2003 12 31 & 14:34:37 & 180 & 2453005.108 & 13.977 & 0.006 \\
2003 12 31 & 14:51:05 & 180 & 2453005.120 & 13.972 & 0.006 \\
2003 12 31 & 15:08:44 & 180 & 2453005.132 & 14.016 & 0.006 \\
2003 12 31 & 15:25:00 & 180 & 2453005.143 & 13.999 & 0.006 \\
2003 12 31 & 15:41:15 & 180 & 2453005.155 & 14.001 & 0.006 \\
2003 12 31 & 15:57:38 & 180 & 2453005.166 & 14.014 & 0.006 \\
2003 12 31 & 16:14:13 & 180 & 2453005.178 & 13.982 & 0.006 \\
2003 12 31 & 16:32:29 & 180 & 2453005.190 & 13.981 & 0.006 \\
2003 12 31 & 16:48:59 & 180 & 2453005.202 & 13.967 & 0.006 \\
2003 12 31 & 17:05:15 & 180 & 2453005.213 & 13.959 & 0.006 \\
2003 12 31 & 17:21:39 & 180 & 2453005.224 & 13.967 & 0.006 \\
2003 12 31 & 17:37:56 & 180 & 2453005.236 & 13.954 & 0.007 \\
2003 12 31 & 17:55:05 & 180 & 2453005.248 & 13.953 & 0.006 \\
2003 12 31 & 18:38:10 & 180 & 2453005.278 & 13.927 & 0.006 \\
2003 12 31 & 19:13:31 & 180 & 2453005.302 & 13.909 & 0.006 \\
2003 12 31 & 19:32:00 & 180 & 2453005.315 & 13.896 & 0.006 \\
2003 12 31 & 19:50:10 & 180 & 2453005.327 & 13.896 & 0.006 \\
2003 12 31 & 20:06:35 & 180 & 2453005.339 & 13.887 & 0.005 \\
2003 12 31 & 20:23:09 & 180 & 2453005.350 & 13.862 & 0.006 \\
2003 12 31 & 20:39:34 & 180 & 2453005.362 & 13.841 & 0.005 \\
2003 12 31 & 20:56:27 & 180 & 2453005.374 & 13.813 & 0.005 \\
2003 12 31 & 21:17:46 & 180 & 2453005.388 & 13.781 & 0.005 \\
2003 12 31 & 21:34:15 & 180 & 2453005.400 & 13.716 & 0.005 \\
2003 12 31 & 21:50:40 & 180 & 2453005.411 & 13.702 & 0.005 \\
2004 01 01 & 12:29:54 & 180 & 2453006.022 & 13.938 & 0.005 \\
2004 01 01 & 12:48:43 & 180 & 2453006.035 & 13.941 & 0.005 \\
2004 01 01 & 13:05:12 & 180 & 2453006.046 & 13.909 & 0.005 \\
2004 01 01 & 13:22:28 & 180 & 2453006.058 & 13.905 & 0.005 \\
2004 01 01 & 13:50:10 & 180 & 2453006.078 & 13.905 & 0.005 \\
2004 01 01 & 14:07:23 & 180 & 2453006.090 & 13.906 & 0.005 \\
2004 01 01 & 14:23:55 & 180 & 2453006.101 & 13.914 & 0.005 \\
2004 01 01 & 14:42:11 & 180 & 2453006.114 & 13.935 & 0.005 \\
2004 01 01 & 15:09:09 & 180 & 2453006.132 & 13.940 & 0.005 \\
2004 01 01 & 15:25:42 & 180 & 2453006.144 & 13.956 & 0.005 \\
2004 01 01 & 15:42:00 & 180 & 2453006.155 & 13.944 & 0.005 \\
2004 01 01 & 16:37:43 & 180 & 2453006.194 & 13.970 & 0.005 \\
2004 01 01 & 16:56:15 & 180 & 2453006.207 & 13.983 & 0.005 \\
2004 01 01 & 17:33:35 & 180 & 2453006.233 & 13.961 & 0.005 \\
2004 01 01 & 17:50:22 & 180 & 2453006.244 & 13.932 & 0.005 \\
2004 01 01 & 18:06:56 & 180 & 2453006.256 & 13.934 & 0.005 \\
2004 01 01 & 18:26:01 & 180 & 2453006.269 & 13.935 & 0.005 \\
2004 01 02 & 12:38:34 & 180 & 2453007.028 & 13.683 & 0.005 \\
2004 01 02 & 12:55:13 & 180 & 2453007.039 & 13.674 & 0.004 \\
2004 01 02 & 13:11:27 & 180 & 2453007.051 & 13.678 & 0.004 \\
2004 01 02 & 13:27:53 & 180 & 2453007.062 & 13.679 & 0.004 \\
2004 01 02 & 13:44:11 & 180 & 2453007.073 & 13.657 & 0.004 \\
2004 01 02 & 14:00:54 & 180 & 2453007.085 & 13.674 & 0.004 \\
2004 01 02 & 14:17:33 & 180 & 2453007.097 & 13.689 & 0.005 \\
2004 01 02 & 14:33:50 & 180 & 2453007.108 & 13.687 & 0.004 \\
2004 01 02 & 14:50:17 & 180 & 2453007.119 & 13.699 & 0.004 \\
2004 01 02 & 15:08:56 & 180 & 2453007.132 & 13.679 & 0.004 \\
2004 01 02 & 15:25:09 & 180 & 2453007.144 & 13.697 & 0.004 \\
2004 01 02 & 15:41:22 & 180 & 2453007.155 & 13.684 & 0.004 \\
2004 01 02 & 15:57:37 & 180 & 2453007.166 & 13.677 & 0.004 \\
2004 01 02 & 16:15:05 & 180 & 2453007.178 & 13.691 & 0.004 \\
2004 01 02 & 16:31:26 & 180 & 2453007.190 & 13.693 & 0.004 \\
2004 01 02 & 16:48:00 & 180 & 2453007.201 & 13.689 & 0.004 \\
2004 01 02 & 17:04:30 & 180 & 2453007.213 & 13.699 & 0.004 \\
2004 01 02 & 17:21:04 & 180 & 2453007.224 & 13.708 & 0.004 \\
2004 01 02 & 17:37:21 & 180 & 2453007.235 & 13.698 & 0.004 \\
2004 01 02 & 17:57:21 & 180 & 2453007.249 & 13.680 & 0.004 \\
2004 01 02 & 18:14:02 & 180 & 2453007.261 & 13.691 & 0.004 \\
2004 01 02 & 18:30:44 & 180 & 2453007.272 & 13.675 & 0.004 \\
2004 01 02 & 18:47:16 & 180 & 2453007.284 & 13.664 & 0.004 \\
2004 01 02 & 19:03:47 & 180 & 2453007.295 & 13.651 & 0.004 \\
2004 01 02 & 19:20:35 & 180 & 2453007.307 & 13.675 & 0.004 \\
2004 01 03 & 15:58:14 & 180 & 2453008.167 & 13.680 & 0.005 \\
2004 01 03 & 16:17:35 & 180 & 2453008.180 & 13.682 & 0.005 \\
2004 01 03 & 16:33:56 & 180 & 2453008.191 & 13.677 & 0.005 \\
2004 01 03 & 16:50:38 & 180 & 2453008.203 & 13.680 & 0.005 \\
2004 01 03 & 17:07:10 & 180 & 2453008.214 & 13.667 & 0.005 \\
2004 01 03 & 17:23:31 & 180 & 2453008.226 & 13.668 & 0.005 \\
2004 01 03 & 17:40:18 & 180 & 2453008.237 & 13.655 & 0.005 \\
2004 01 03 & 17:57:35 & 180 & 2453008.249 & 13.652 & 0.005 \\
2004 01 03 & 18:14:09 & 180 & 2453008.261 & 13.624 & 0.005 \\
2004 01 03 & 18:30:26 & 180 & 2453008.272 & 13.657 & 0.005 \\
2004 01 03 & 18:47:04 & 180 & 2453008.284 & 13.642 & 0.005 \\
2004 01 03 & 19:03:35 & 180 & 2453008.295 & 13.627 & 0.005 \\
2004 01 03 & 19:20:29 & 180 & 2453008.307 & 13.629 & 0.005 \\
2004 01 03 & 19:36:57 & 180 & 2453008.318 & 13.636 & 0.005 \\
2004 01 03 & 19:53:13 & 180 & 2453008.330 & 13.640 & 0.005 \\
2004 01 03 & 20:10:24 & 180 & 2453008.342 & 13.644 & 0.005 \\
2004 01 03 & 20:27:21 & 180 & 2453008.353 & 13.643 & 0.005 \\
2004 01 03 & 20:48:33 & 180 & 2453008.368 & 13.656 & 0.004 \\
2004 01 04 & 12:06:32 & 180 & 2453009.006 & 13.709 & 0.006 \\
2004 01 04 & 12:23:00 & 180 & 2453009.017 & 13.718 & 0.005 \\
2004 01 04 & 12:43:48 & 180 & 2453009.032 & 13.712 & 0.005 \\
2004 01 04 & 13:01:58 & 180 & 2453009.044 & 13.723 & 0.005 \\
2004 01 04 & 13:18:13 & 180 & 2453009.055 & 13.706 & 0.005 \\
2004 01 04 & 15:52:13 & 180 & 2453009.162 & 13.658 & 0.005 \\
2004 01 04 & 16:09:33 & 180 & 2453009.174 & 13.640 & 0.005 \\
2004 01 04 & 16:25:51 & 180 & 2453009.186 & 13.627 & 0.005 \\
2004 01 04 & 16:42:19 & 180 & 2453009.197 & 13.636 & 0.005 \\
2004 01 04 & 16:58:33 & 180 & 2453009.208 & 13.610 & 0.005 \\
2004 01 04 & 17:16:36 & 180 & 2453009.221 & 13.577 & 0.005 \\
2004 01 04 & 17:33:17 & 180 & 2453009.233 & 13.582 & 0.005 \\
2004 01 04 & 17:50:25 & 180 & 2453009.244 & 13.599 & 0.005 \\
2004 01 04 & 18:07:07 & 180 & 2453009.256 & 13.577 & 0.005 \\
2004 01 04 & 18:23:25 & 180 & 2453009.267 & 13.573 & 0.005 \\
2004 01 04 & 18:41:46 & 180 & 2453009.280 & 13.568 & 0.005 \\
2004 01 04 & 18:58:27 & 180 & 2453009.292 & 13.549 & 0.005 \\
2004 01 04 & 19:15:08 & 180 & 2453009.303 & 13.514 & 0.004 \\
2004 01 04 & 19:31:45 & 180 & 2453009.315 & 13.547 & 0.005 \\
2004 01 04 & 19:48:09 & 180 & 2453009.326 & 13.529 & 0.004 \\
2004 01 04 & 20:06:37 & 180 & 2453009.339 & 13.515 & 0.005 \\
2004 01 04 & 20:23:15 & 180 & 2453009.350 & 13.506 & 0.004 \\
2004 01 04 & 20:39:46 & 180 & 2453009.362 & 13.488 & 0.004 \\
2004 01 04 & 20:56:28 & 180 & 2453009.374 & 13.489 & 0.004 \\
2004 01 04 & 21:13:05 & 180 & 2453009.385 & 13.466 & 0.004 \\
2004 01 04 & 21:30:37 & 180 & 2453009.397 & 13.453 & 0.004 \\
2004 01 05 & 13:22:30 & 180 & 2453010.058 & 13.514 & 0.005 \\
2004 01 05 & 13:53:18 & 180 & 2453010.080 & 13.464 & 0.005 \\
2004 01 05 & 14:09:54 & 180 & 2453010.091 & 13.463 & 0.005 \\
2004 01 05 & 14:26:34 & 180 & 2453010.103 & 13.488 & 0.007 \\
\enddata
\end{deluxetable}

\end{document}